\documentclass[pra,reprint,amsmath,amssymb,aps,superscriptaddress]{revtex4-2}
\usepackage{graphicx}
\usepackage{hyperref}
\usepackage{color}
\usepackage{makecell}
\usepackage{listings}

\begin{document}
\title{No More Hooks in the Surface Code: Distance-Preserving Syndrome Extraction for Arbitrary Layouts at Minimum Depth}

\author{Yuga Hirai}
\email{yuga917@keio.jp}
\affiliation{Department of Applied Physics and Physico-Informatics, Keio University, Kanagawa, Japan}

\author{Shota Ikari}
\email{shota-ikari@g.ecc.u-tokyo.ac.jp}
\affiliation{Graduate School of Information Science and Technology, The University of Tokyo, Tokyo, Japan}

\author{Yosuke Ueno}
\email{yosuke.ueno@riken.jp}
\affiliation{Center for Quantum Computing, RIKEN, Saitama, Japan}
\affiliation{Graduate School of Information Science and Technology, The University of Tokyo, Tokyo, Japan}

\author{Yasunari Suzuki}
\email{yasunari.suzuki@riken.jp}
\affiliation{Center for Quantum Computing, RIKEN, Saitama, Japan}

\date{\today}

\begin{abstract}
Hook errors are a major challenge in implementing logical operations with the surface code, because they can reduce the fault distance below the code distance. This motivates syndrome-extraction circuits that suppress hook-error effects for the stabilizer layouts that appear during logical operations. However, the existing methods either increase circuit depth or require simultaneous execution of measurements and CNOT gates, both of which introduce additional overheads and degrade the threshold. We propose the ZX interleaving syndrome extraction, which preserves the full fault distance $d$ for any surface-code layout with regular stabilizer tiles at minimum depth, i.e., four layers of CNOT gates, without requiring additional circuit depth or simultaneous execution of measurements and CNOT gates. The key idea is to interleave the Z and X stabilizer tiles so that hook-error edges in the decoding graph are shortened and effectively eliminated. Numerical simulations under uniform depolarizing noise for memory and lattice-surgery experiments confirm that the proposed method achieves a full fault distance of $d$, whereas the best existing minimum-depth approach achieves $d-1$. Since the full fault distance is achievable for any regular tiling layout of the surface code, the proposed method may serve as an indispensable technique for practical fault-tolerant quantum computation.
\end{abstract}
\maketitle

\section{Introduction}
Recent experimental breakthroughs have demonstrated quantum error correction on physical devices~\cite{google2025quantum,google2023suppressing,lacroix2025scaling,bluvstein2024logical,rosenfeld2025magic}, marking a significant step toward practical fault-tolerant quantum computation~(FTQC). Nevertheless, substantial improvements in the performance of FTQC are still required.

Currently, surface code is the most practical quantum error-correcting~(QEC) code because it requires only nearest-neighbor interactions for syndrome extraction and can realize a universal logical gate set comprising H, S, CNOT, and T gates through lattice surgery~\cite{horsman2012surface,fowler2018low}, code deformation~\cite{bombin2009quantum, chamberland2022universal,gidney2024inplace}, magic state cultivation~\cite{gidney2024magic,hirano2025efficient}, or magic state distillation~\cite{bravyi2005universal}. Also, it can be efficiently decoded by the minimum-weight perfect matching algorithm~\cite{higgott2022pymatching}. However, performing arbitrary logical operations in the surface code on real devices remains challenging. 

A major obstacle is hook error. 
Hook error is a single-qubit Pauli error propagating to multiple physical qubits during syndrome extraction, which can degrade the fault distance of the code. 
In general, hook errors are unavoidable when employing standard minimum-depth syndrome-extraction circuits consisting of four layers of CNOT gates \cite{tomita2014low,heim2016optimal}. 
While a minimum-depth syndrome-extraction circuit with a full fault distance is known for memory experiments, such syndrome-extraction circuits for stabilizer layouts that appear in logical operations (e.g., lattice surgery) have yet to be proposed. 

Existing methods typically circumvent hook errors at the cost of additional overheads, such as additional circuit depth or the simultaneous execution of measurements and CNOT gates~\cite{kishony2026surface,litinski2018lattice}. 
While adding extra depth to the syndrome-extraction circuit provides greater flexibility in scheduling CNOT gates, it also increases the total execution time for syndrome extraction. 
The increased execution time per syndrome extraction results in higher effective physical error rates per cycle in the code, thereby degrading the threshold, the physical error rate below which the logical error rates are suppressed with increasing code distance. 
Similarly, the simultaneous execution of measurements and CNOT gates poses practical challenges because these operations typically have different durations~\cite{google2023suppressing,google2025quantum}; the faster operation is forced to idle, which again increases the effective cycle time. 
To summarize, the limitations of existing methods are:
\begin{itemize}
\item Increasing circuit depth.
\item Requiring simultaneous measurement and CNOT execution.
\item Non-uniformity of CNOT ordering across the entire surface code.
\end{itemize}
These drawbacks motivate our search for a new approach.

In this work, we propose a new method that resolves all the aforementioned issues. 
The key idea is to exploit the syndrome-extraction scheme originally proposed for the dynamic surface code~\cite{mcewen2023relaxing}, where Z-stabilizer tiles move toward the Z boundary and X-stabilizer tiles toward the X boundary. 
As we detailed later, these movements effectively eliminate hook errors by shortening the corresponding error-propagation edges within the decoding graph. 
However, this original scheme is specific to surface-code-memory experiments and cannot be directly applied to general stabilizer layouts such as those used in lattice surgery.
This limitation arises because general layouts introduce scenarios where Z and X tiles must move toward boundaries of the opposite Pauli type. 
To overcome this, we introduce a scheme that accommodates such cases at minimum circuit depth by only deploying additional measurement qubits along the boundaries. 

We numerically evaluate our method alongside existing methods through memory experiments and logical Pauli-XX measurement involving lattice surgery. 
In both scenarios, our method is the only one that preserves the full fault distance at minimum depth. 
In the low physical error rate regime, the logical error rates of our method scale nearly identically to those of the full-fault-distance implementations in memory experiments, and it achieves the lowest logical error rates among minimum-depth methods in logical Pauli-XX measurements.
We note a slight trade-off: due to the introduction of additional measurement qubits and a more complex CNOT schedule, our method exhibits a marginally lower threshold compared to existing methods, making it less advantageous when operating near the threshold. 
Nevertheless, to the best of our knowledge, this is the first method to preserve the full fault distance at minimum depth for any surface code layout with regular stabilizer tiles.

Additionally, our method requires measurement qubits to couple with only three neighboring data qubits. 
As demonstrated in Ref.~\cite{eickbusch2025demonstration}, reducing connectivity can suppress circuit-level physical error rates, which may well offset the slight degradation in the threshold of our method. 
Because our syndrome-extraction circuit keeps the full fault distance, achieves the minimum CNOT depth, and realizes any layout required by lattice surgery, patch movement, or patch rotation, it is an indispensable technique for practical FTQC.

The remainder of this paper is organized as follows. 
Section~\ref{sec:preliminaries} reviews the preliminaries, including the rotated surface code, hook errors, and utilities to discuss complicated syndrome-extraction circuits such as decoding graphs and detector diagrams. 
Section~\ref{sec:motivation} summarizes a problem of hook errors and reviews the existing methods for distance-preserving syndrome extraction. 
Section~\ref{sec:zx_interleaving} presents the proposed ZX interleaving syndrome extraction and explains how it eliminates hook errors for arbitrary layouts at minimum depth. 
Section~\ref{sec:numerical} reports numerical results for memory experiments and logical Pauli-XX measurement experiments under uniform depolarizing noise. 
Finally, Section~\ref{sec:conclusion} concludes the paper and discusses directions for future work.

\section{Preliminaries}
\label{sec:preliminaries}
In this section, we review the fundamental concepts required to understand this paper, including the rotated surface code, logical operations, hook errors, decoding graph, and detector diagrams.
For more details on preliminary topics of QEC codes and their fault distances, please refer to Refs.~\cite{mcewen2023relaxing,gidney2021stim}.

\subsection{Surface Code and Logical Operations}
Quantum error correction is an essential technique for realizing practical quantum computation. It can protect quantum information by encoding it into a code space defined as the simultaneous +1 eigenspace of a commutative set of Pauli operators called stabilizer generators. 
Surface codes~\cite{kitaev1997quantum} are known as one of the most promising QEC codes, because they can implement universal logical operations only with nearest-neighbor interactions of physical qubits allocated on two-dimensional~(2D) grids~\cite{fowler2018low,beverland2022assessing}.

\begin{figure}[t]
  \centering
  \includegraphics[width=0.9\columnwidth]{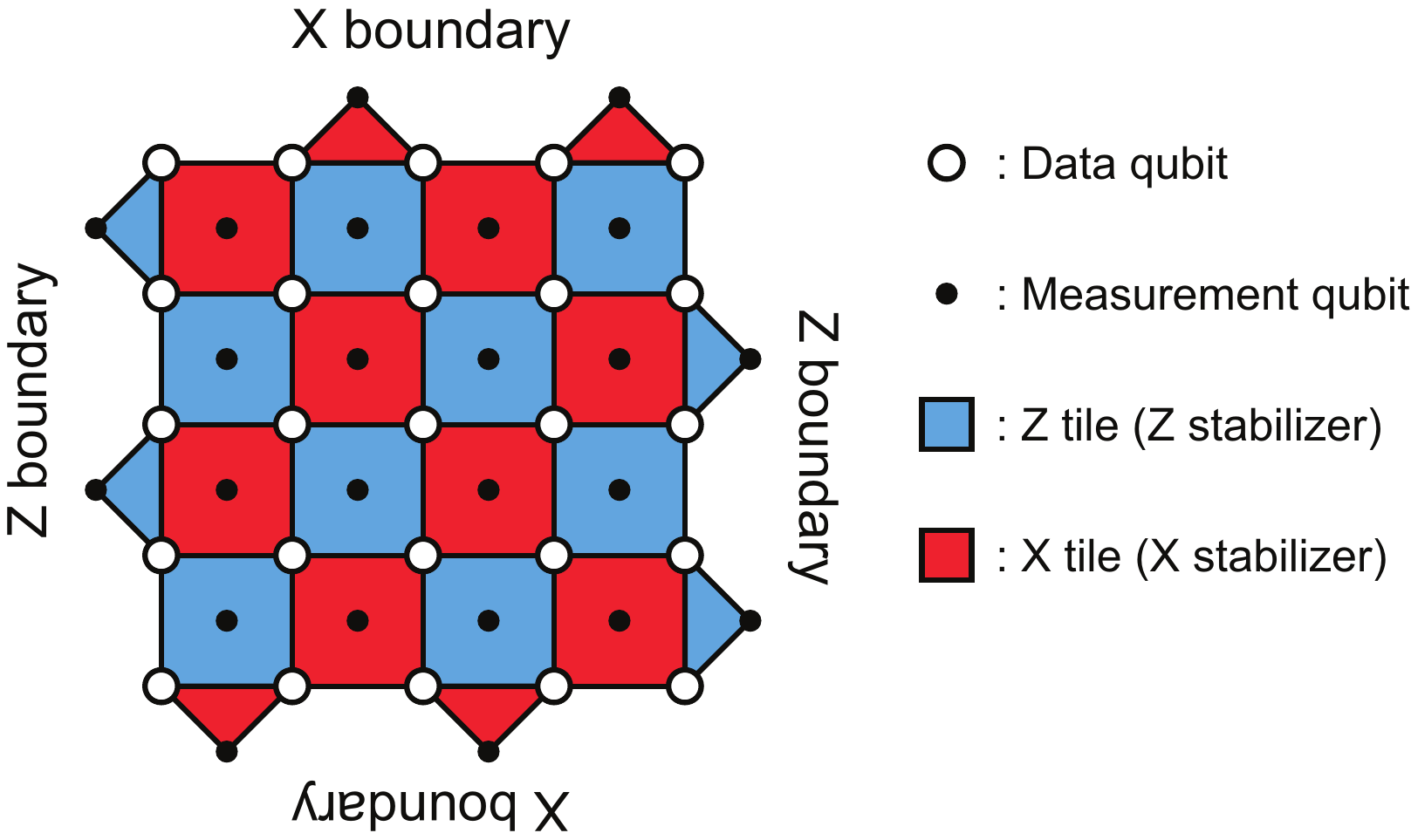}
  \caption{Rotated surface code with code distance $d=5$.}
  \label{fig:surface_code}
\end{figure}

This paper focuses on QEC with rotated surface codes, whose qubit allocations and stabilizer generators are shown in Fig.~\ref{fig:surface_code}. The white and black circles represent data and measurement qubits, respectively. Data qubits store the quantum information, while measurement qubits are used to extract the syndrome from the data qubits. The blue and red faces represent the Z and X stabilizers acting on the data qubits at the corners, respectively. Surface codes repeatedly run a syndrome extraction circuit to continuously estimate errors on physical qubits.
Throughout this paper, we use the term \textit{tiles} to refer to these tile-like stabilizers in the surface code, and the term \textit{patch} to refer to the whole code block. Boundaries terminated by Z~(X) tiles are called Z~(X) boundaries. The logical Z~(X) operator is supported on data qubits connecting the two opposing Z~(X) boundaries. The code distance of the surface code is defined by the side length of the lattice; for example, the code distance in Fig.~\ref{fig:surface_code} is $d=5$.

\begin{figure}[t]
  \centering
  \includegraphics[width=0.9\columnwidth]{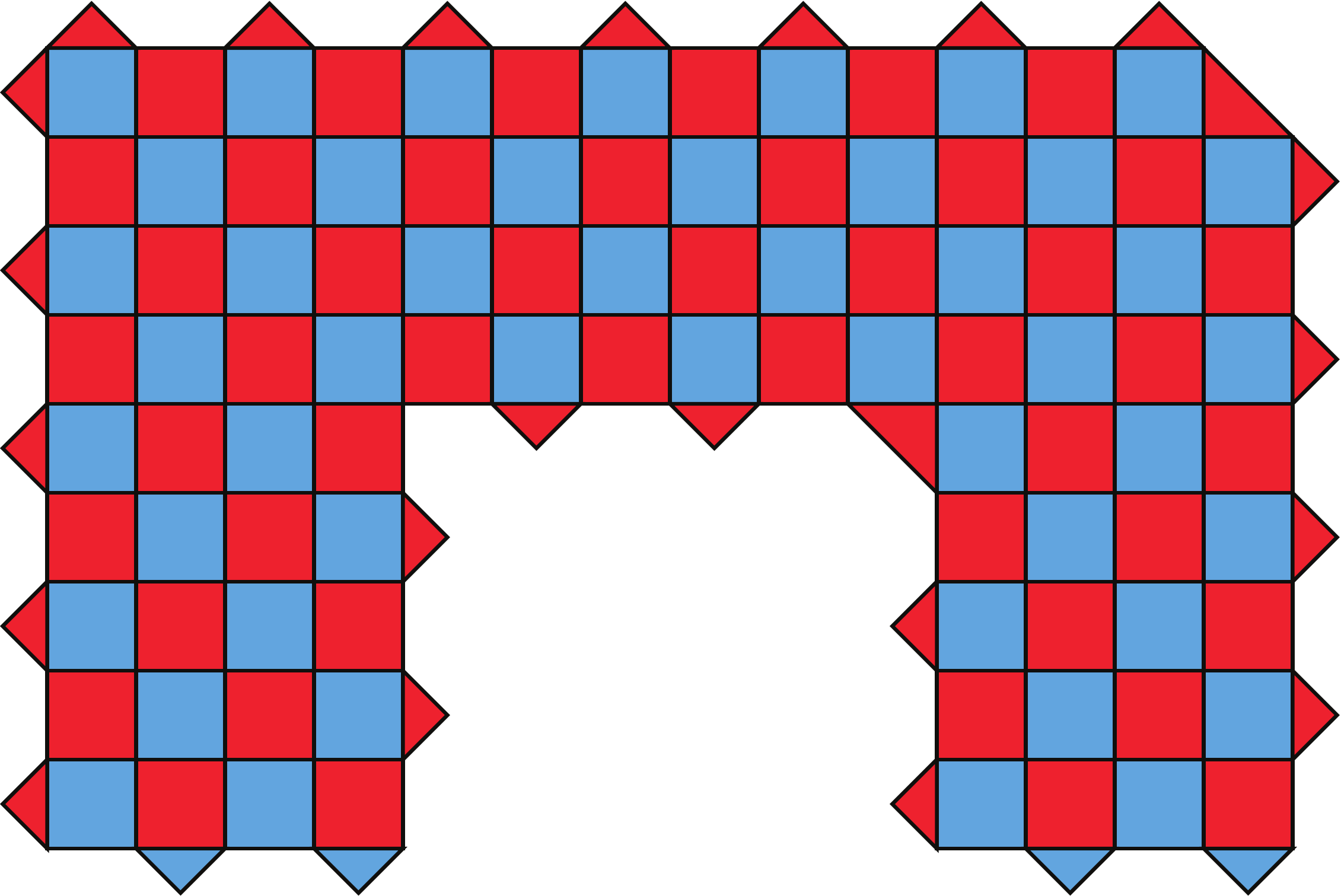}
  \caption{Surface code layout of the logical Pauli-XX measurement using lattice surgery.}
  \label{fig:XX_measurement}
\end{figure}

In the surface code, several logical Clifford operations are realized through lattice surgery and code deformation~\cite{bombin2023logical,chamberland2022universal,gidney2024inplace,horsman2012surface,bravyi2005universal,gidney2024magic,hirano2025efficient}. They are techniques to perform logical operations on encoded information in a fault-tolerant manner by changing a set of stabilizer generators. For example, when we perform logical Pauli-XX measurements on two distant patches of surface codes, we perform stabilizer measurements with a layout where their boundaries are connected, as shown in Fig.~\ref{fig:XX_measurement}, several times, and then they are separated to the original form. Similarly, by modifying the tiles and boundaries, we can perform a patch rotation after a transversal Hadamard gate~\cite{litinski2019game,fowler2018low}, a logical S-gate~\cite{bombin2023logical} with twist defects, and a patch movement~\cite{litinski2019game}.

\subsection{Syndrome Extraction and Hook Errors}
\begin{figure}[tb]
    \centering
    \includegraphics[width=0.4\textwidth]{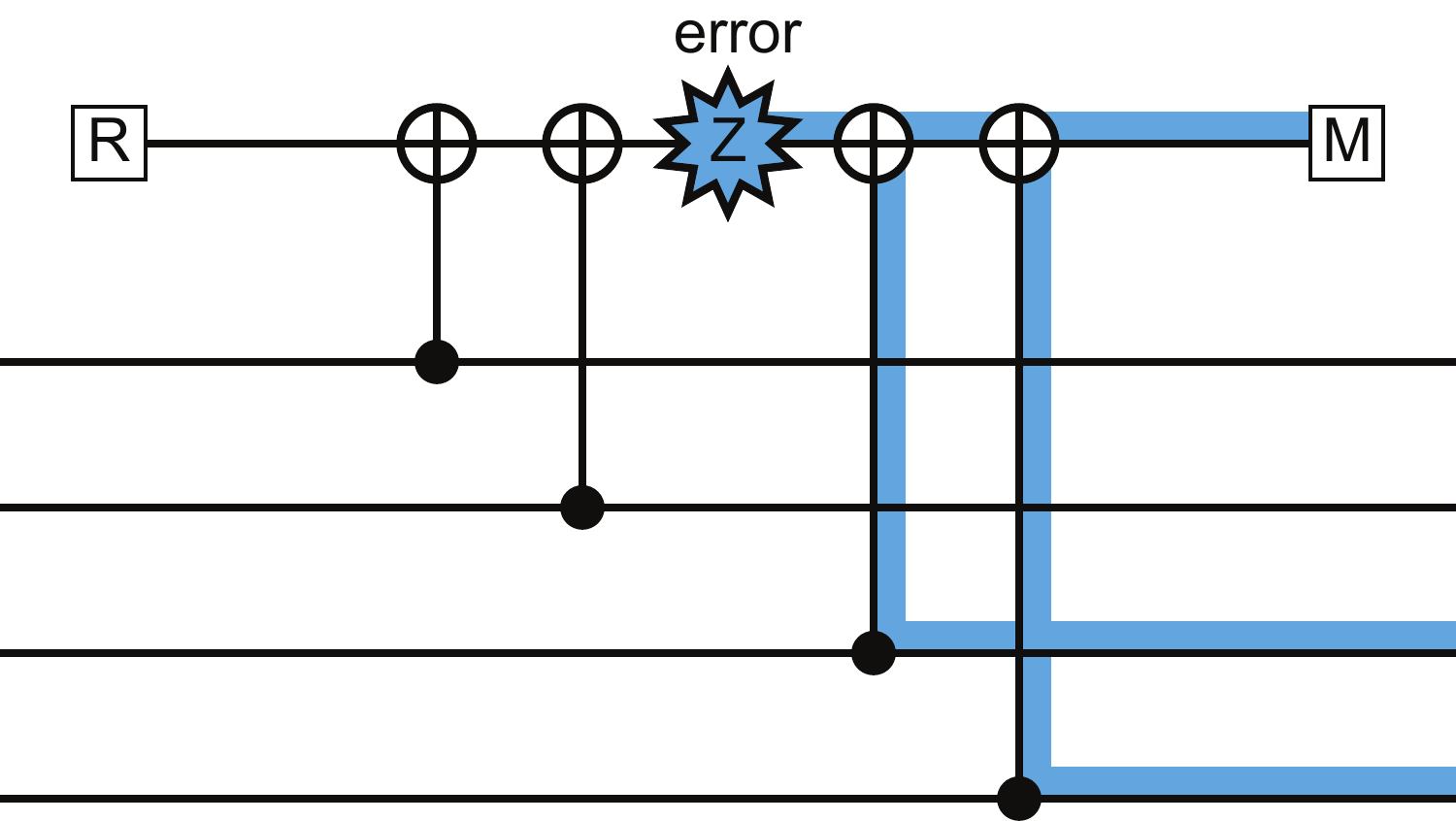}
    \caption{Syndrome extraction circuit. The Z error is coupled to the two data qubits via the CNOT gates shown in blue.}
    \label{fig:hook_error}
\end{figure}

Surface codes with distance $d$ can detect any Pauli error acting on fewer than $d$ physical qubits if we can assume errors occur between rounds of stabilizer measurements. However, physical errors can occur during syndrome extraction in practice, and a single-qubit Pauli error can propagate to multiple physical qubits. Such an error is called a hook error, and hook errors may reduce the effective code distance, which is the minimum number of physical errors in the circuit required to cause logical errors, known as the \textit{fault distance}. An example of a hook error is shown in Fig.~\ref{fig:hook_error}.
The figure depicts a syndrome-extraction circuit for the Z stabilizers, where R and M denote a reset and a measurement operation, respectively. In this example, an error occurs on the measurement qubit after the first two CNOT operations and then propagates to two data qubits, because a CNOT gate copies a Z gate from the target to the control qubits. Similarly, in the circuit for the X stabilizers, X errors can propagate in the same manner. If hook errors span the support of a logical operator, the fault distance becomes smaller than the code distance. Therefore, the ordering of CNOT gates must be carefully designed to preserve the fault distance.

\begin{figure}[tb]
    \centering
    \includegraphics[width=0.2\textwidth]{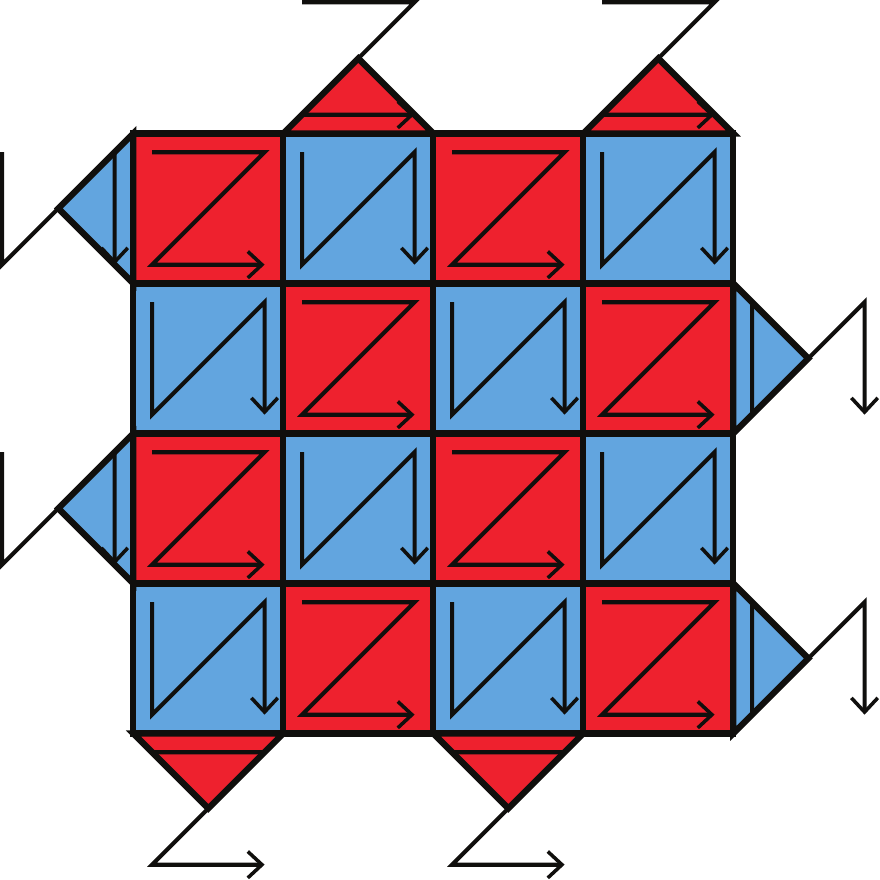}
    \caption{An order of the CNOT gates in the syndrome extraction circuit that preserves the fault distance in the memory layout.}
    \label{fig:surface_order}
\end{figure}

Fortunately, a CNOT ordering that preserves the fault distance of the memory layout (Fig.~\ref{fig:surface_code}) with minimum depth, i.e., four layers of CNOT gates, is known as shown in Fig.~\ref{fig:surface_order}~\cite{tomita2014low,heim2016optimal}. Each arrow indicates the CNOT ordering within a tile, where the tail corresponds to the first gate and the head to the last. The N- and Z-shaped orderings are used for the Z and X stabilizers, respectively. With this ordering, hook errors propagate perpendicular to the logical operators; thus, although hook errors still occur, they do not reduce the fault distance. However, it is not trivial to design a short CNOT ordering that preserves the fault distance during logical operations, such as a layout shown in Fig.~\ref{fig:XX_measurement}. This is a problem this paper addresses.

\subsection{Decoding Graph and Detector Diagrams}
A \textit{decoding graph} and \textit{detector diagram}, proposed in the framework of Stim~\cite{gidney2021stim,mcewen2023relaxing}, are powerful utilities to monitor the error propagation and fault distance during complicated syndrome-extraction circuits. This section provides a brief explanation of these models. Note that while these concepts can be applied to any QEC code, our explanation focuses on the case of the rotated surface code.

\begin{figure}[t]
    \centering
    \includegraphics[width=0.2\textwidth]{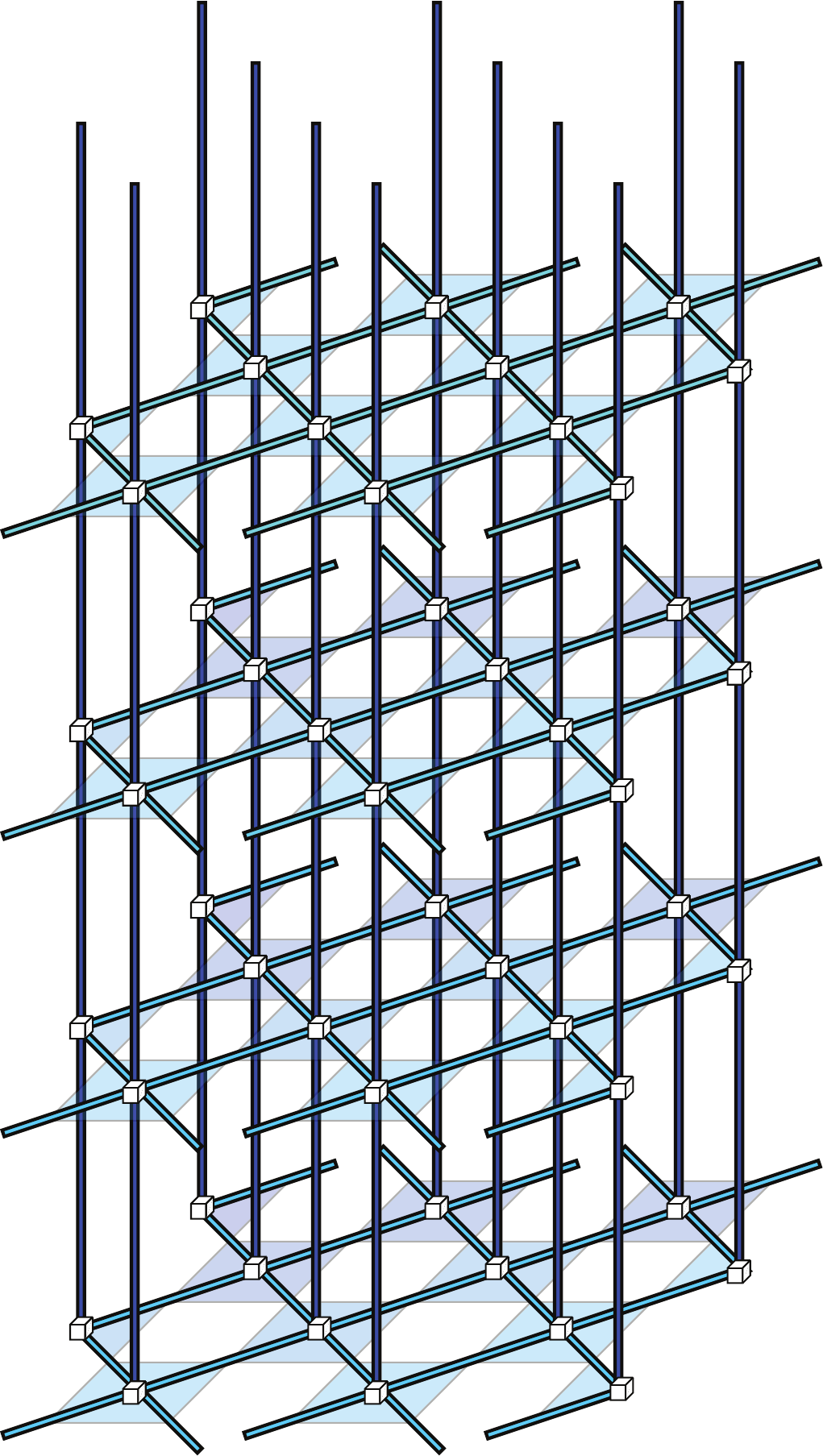}
    \caption{Decoding graph for the Z detectors in the surface code.}
    \label{fig:dec_graph_surface}
\end{figure}
\subsubsection{Decoding Graph}
The mechanism of error detection and estimation can be formalized as follows. We refer to the values obtained from Pauli measurements during syndrome extraction as {\it records}. We define {\it logical observables} as XORs of records to represent values corresponding to logical Pauli-measurement outcomes. Also, we define \textit{detectors} as XORs of records, which must always be zero if there is no error.
Supposing that we enumerate all the possible Pauli-error events on a syndrome-extraction circuit, we can efficiently track what detectors and logical observables are flipped by each error event using the stabilizer formalism. Thus, the error estimation task can be formalized as the estimation of logical observable flips from detector values. 

A Pauli error that flips at most two detectors is called {\it graphlike}. For most logical operations of surface codes, we can design detectors so that every error becomes graphlike or can be decomposed as a product of graphlike errors. In such cases, we can define a decoding graph as a graph in which detectors correspond to nodes. A graphlike error that flips two detectors is mapped to an edge connecting the corresponding nodes. A graphlike error that flips one detector becomes an error connecting a node to a Z or X boundary, depending on its error type. Then, we can upper-bound the fault distance as the number of edges required to connect two opposing Z or X boundaries, and this bound becomes tight when we can ignore errors that are decomposed to a product of graphlike errors.

Figure~\ref{fig:dec_graph_surface} shows the decoding graph of Z detectors for the surface codes, where edges corresponding to hook errors are omitted. Here, the cubes represent detectors. The light blue edges are called spacelike edges and correspond to data-qubit errors, and the dark blue edges are called timelike edges and correspond to measurement-qubit errors. In this figure, the front- and back-sides correspond to opposing Z boundaries, and we need five edges to connect them. These discussions are analogous to the X-stabilizer decoding graph.

\begin{figure}[t]
    \centering
    \includegraphics[width=0.5\textwidth]{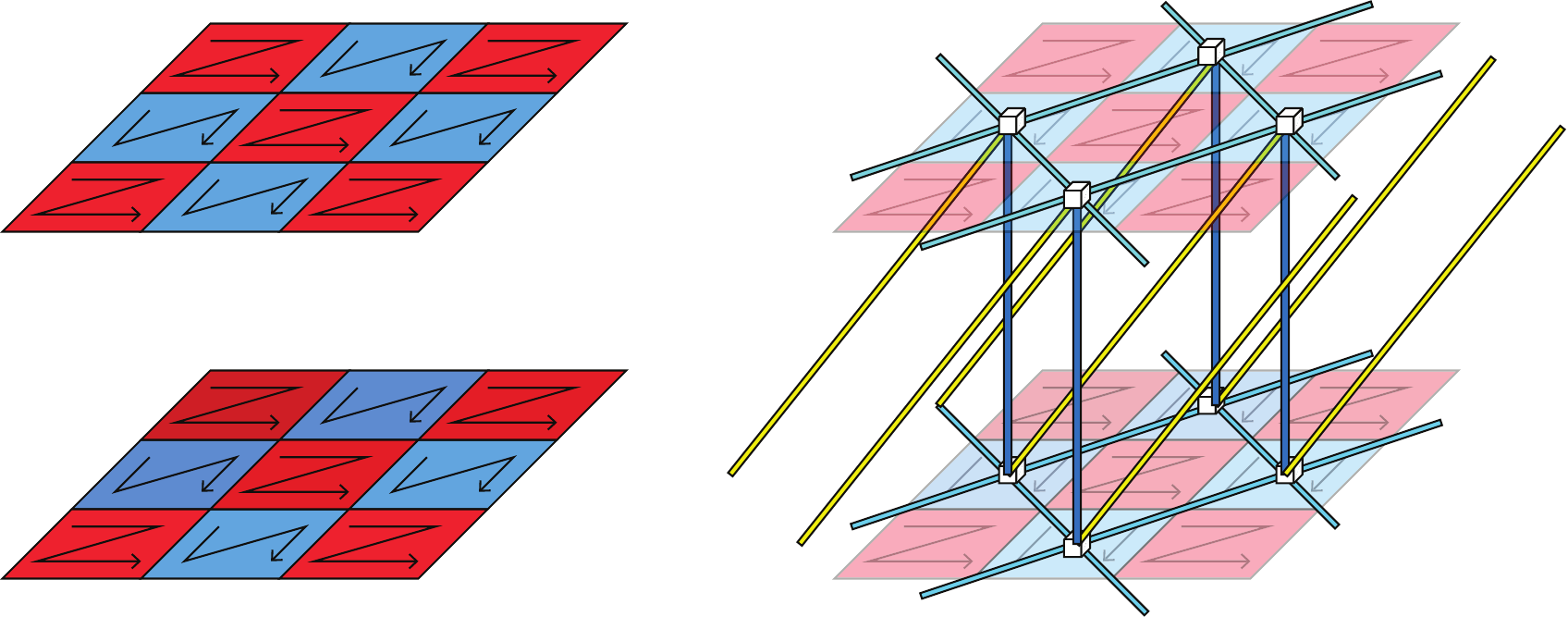}
    \caption{The edges corresponding to hook errors in the Z-stabilizer decoding graph.}
    \label{fig:hook_errors_in_dec_graph}
\end{figure}

Let us now examine the bulk of the surface code more closely to understand how hook errors are expressed in the decoding graph. The edges corresponding to hook errors are shown by the yellow edges in Fig.~\ref{fig:hook_errors_in_dec_graph}. As can be seen from the figure, hook errors connect two detectors separated by a tile, effectively creating shortcut paths in the decoding graph. In this figure, the hook-error edges span from left to right. While they may reduce the fault distance in general, this shortcut does not reduce it in the case of Fig.~\ref{fig:dec_graph_surface}, since the support of the logical Z operator spans from the front to the back. 

\begin{figure}[t]
    \centering
    \includegraphics[width=0.4\textwidth]{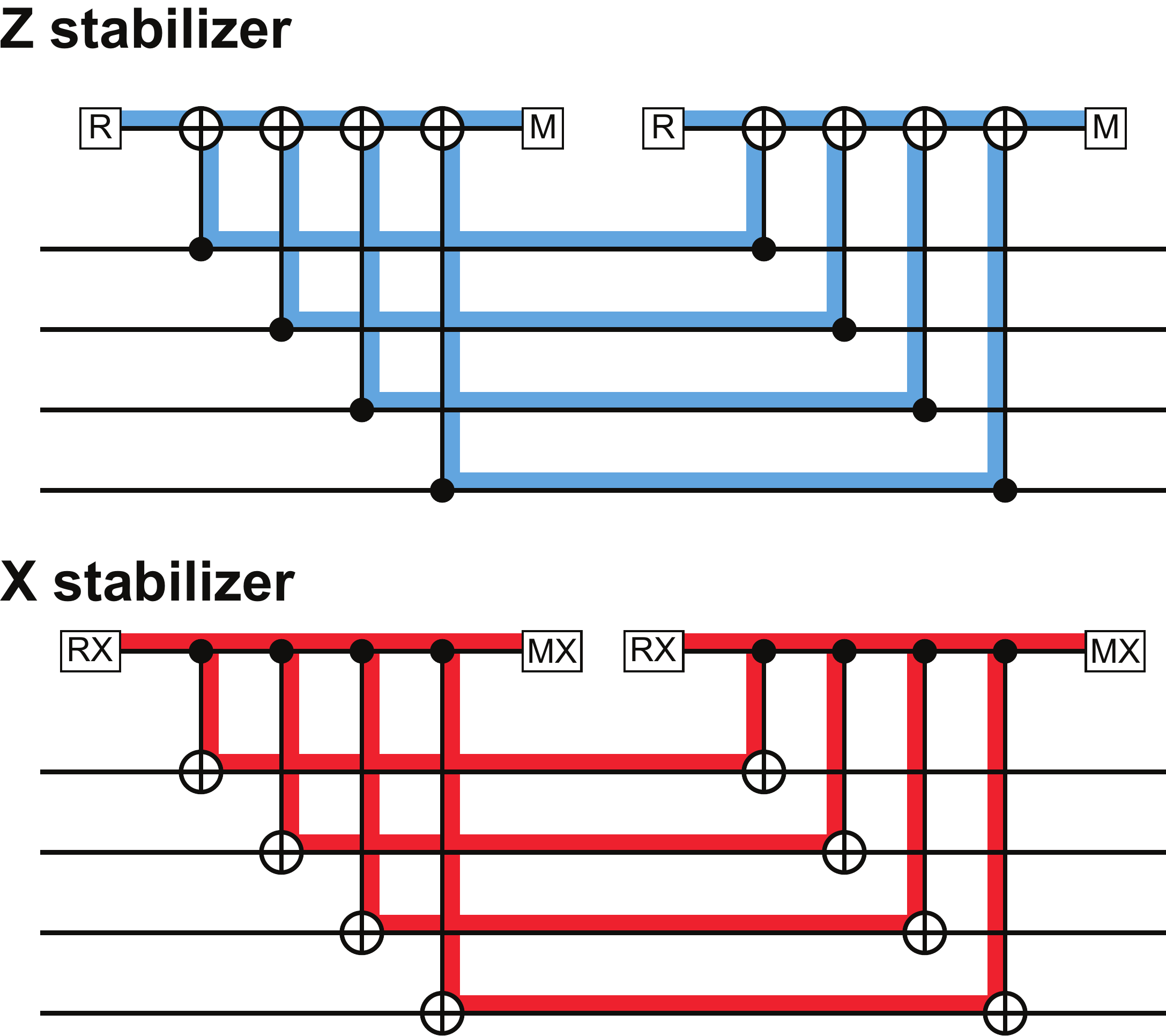}
    \caption{An example of detecting regions.}
    \label{fig:detecting_region}
\end{figure}
\subsubsection{Detector Diagram}
\begin{figure}[t]
    \centering
    \includegraphics[width=0.5\textwidth]{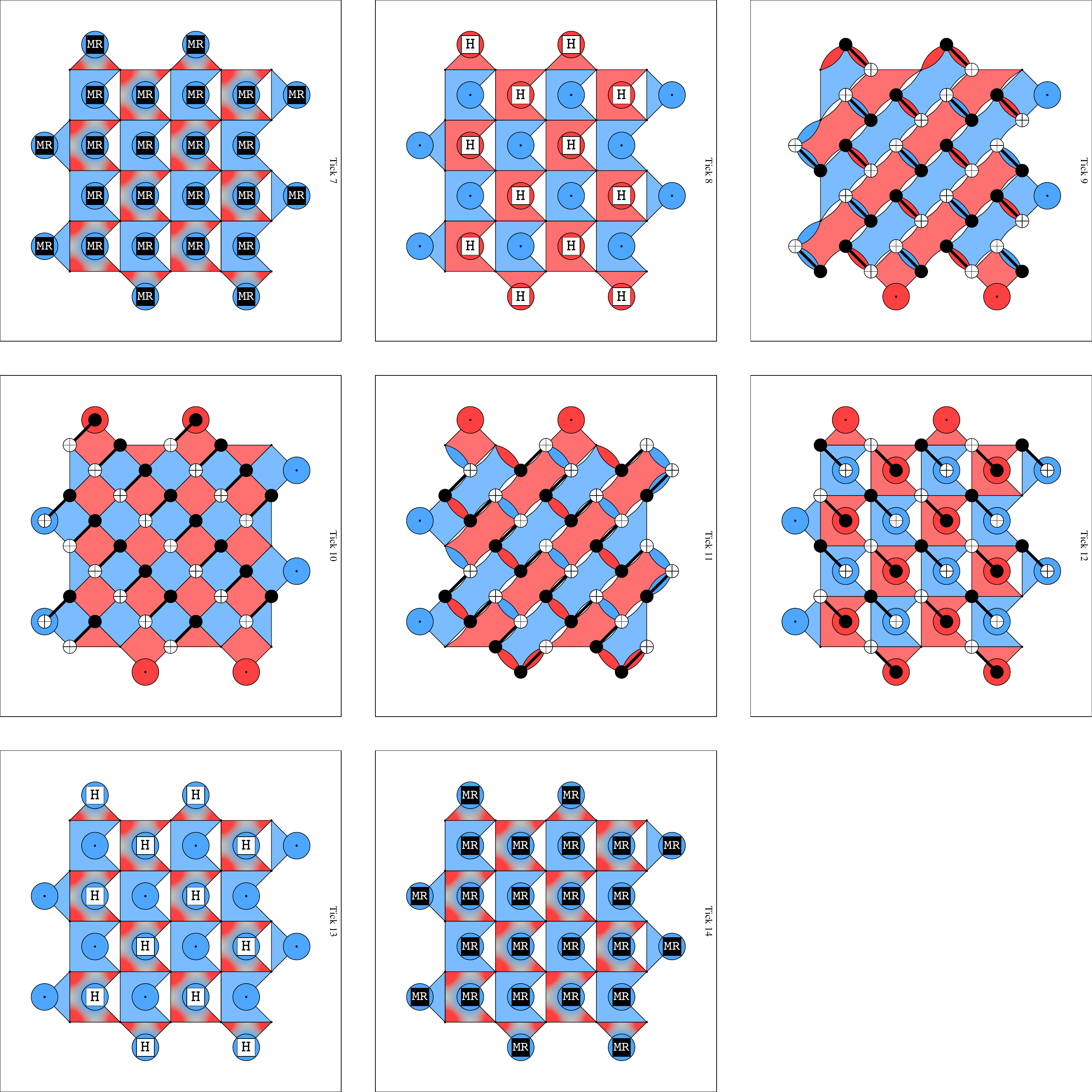}
    \caption{Detector diagram of a single round of the syndrome extraction circuit for the surface code.}
    \label{fig:det_diagram_surface}
\end{figure}

A detector diagram is a useful visualization to check that all the relevant physical qubits are monitored by detectors during syndrome-extraction circuits. Detector diagrams are defined via the concept of a {\it detecting region}, which is a region of a quantum circuit where errors flip a detector. Figure~\ref{fig:detecting_region} shows an example of the detection region for a detector defined as an XOR of two consecutive record values of the measurement qubits.
Then, detector diagrams are defined as a time-slice sequence of the detection regions. For example, Fig~\ref{fig:det_diagram_surface} illustrates how detecting regions are modified during a single round of syndrome extraction. The blue and red region detects X and Z errors, respectively. If an error occurs within a region, the opposite type of detector is flipped.

\begin{figure}[tb]
    \centering
    \includegraphics[width=0.4\textwidth]{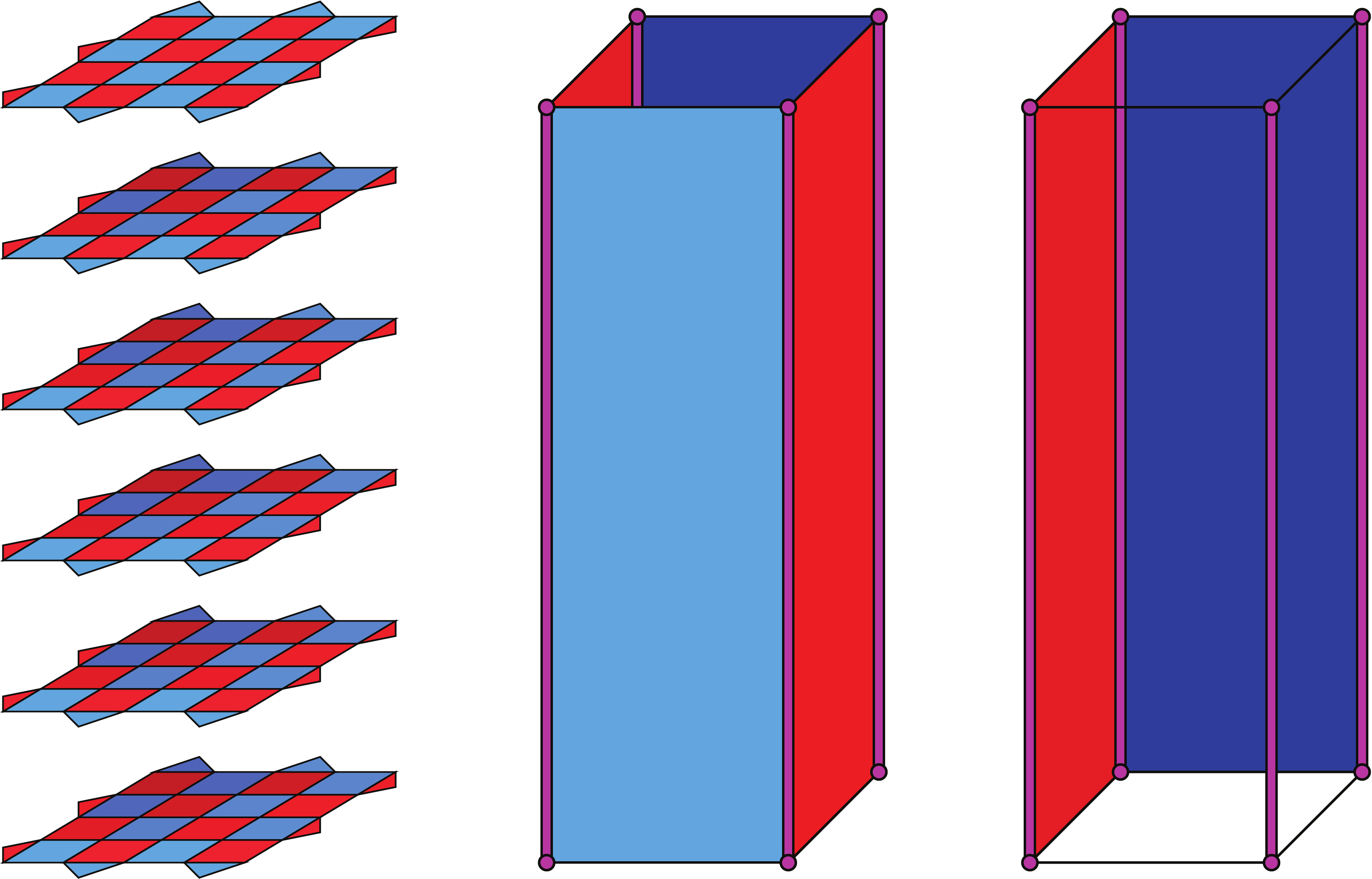}
    \caption{The memory experiments for the surface code. (Left) Surface codes stacked along the time axis. (Middle) A 3D spacetime diagram of the
    surface code. (Right) The same 3D diagram with transparent front walls. Time goes upward.}
    \label{fig:def_3D_diagram}
\end{figure}

\begin{figure}[t]
    \centering
    \includegraphics[width=0.5\textwidth]{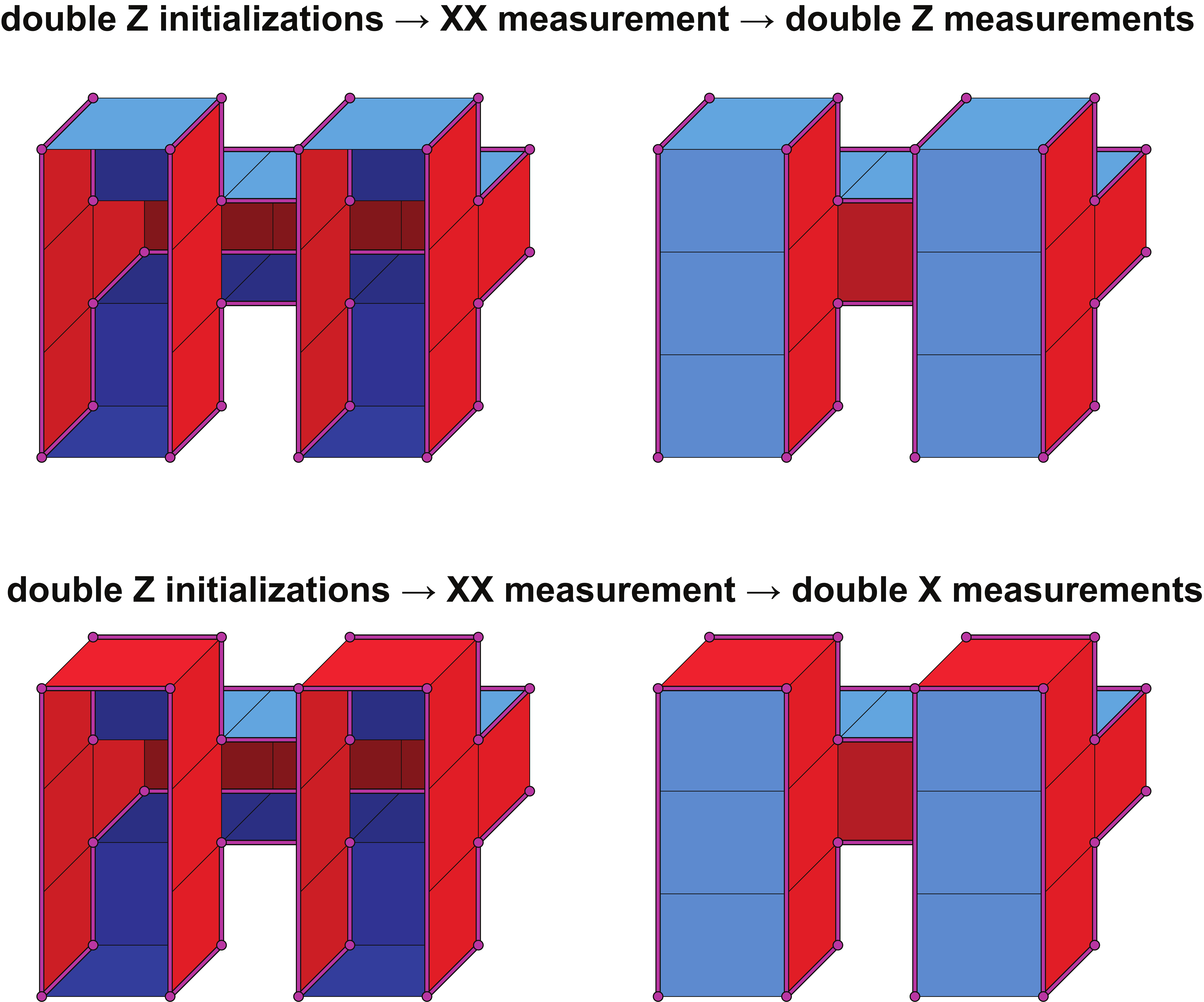}
    \caption{3D spacetime diagram of the logical Pauli-XX measurement for the surface code.}
    \label{fig:XX_meas}
\end{figure}

We can also use detecting regions to visualize a process of logical operations as a 3D diagram, where the Z-axis corresponds to the time dimension~\cite{gidney2024inplace}. A 3D diagram for the memory experiment is shown in Fig.~\ref{fig:def_3D_diagram}. The left panel shows surface codes stacked along the time axis. The middle panel shows the corresponding 3D spacetime diagram, where the blue and red walls represent the Z and X boundaries extended along the time axis, and the purple lines represent twist defects at the intersections of Z and X boundaries. The right panel shows the same 3D diagram with the front walls made transparent. Similarly, we can also illustrate the 3D diagram of the logical Pauli XX-measurement as shown in Fig.~\ref{fig:XX_meas}.

Here, logical Pauli operators are defined analogously to the 2D case: an error chain that connects the two distinct Z~(X) walls works as a logical Pauli Z~(X) error. Thus, to preserve the fault distance, distinct walls of the same color must be separated by at least the code distance, where the distance is measured as the number of edges in the decoding graph.

\section{Motivation: Distance-Preserving Syndrome Extraction Circuits for arbitrary tile layouts}
\label{sec:motivation}

\subsection{Problem setting}
\begin{figure}[tb] 
\centering
\includegraphics[width=0.9\columnwidth]{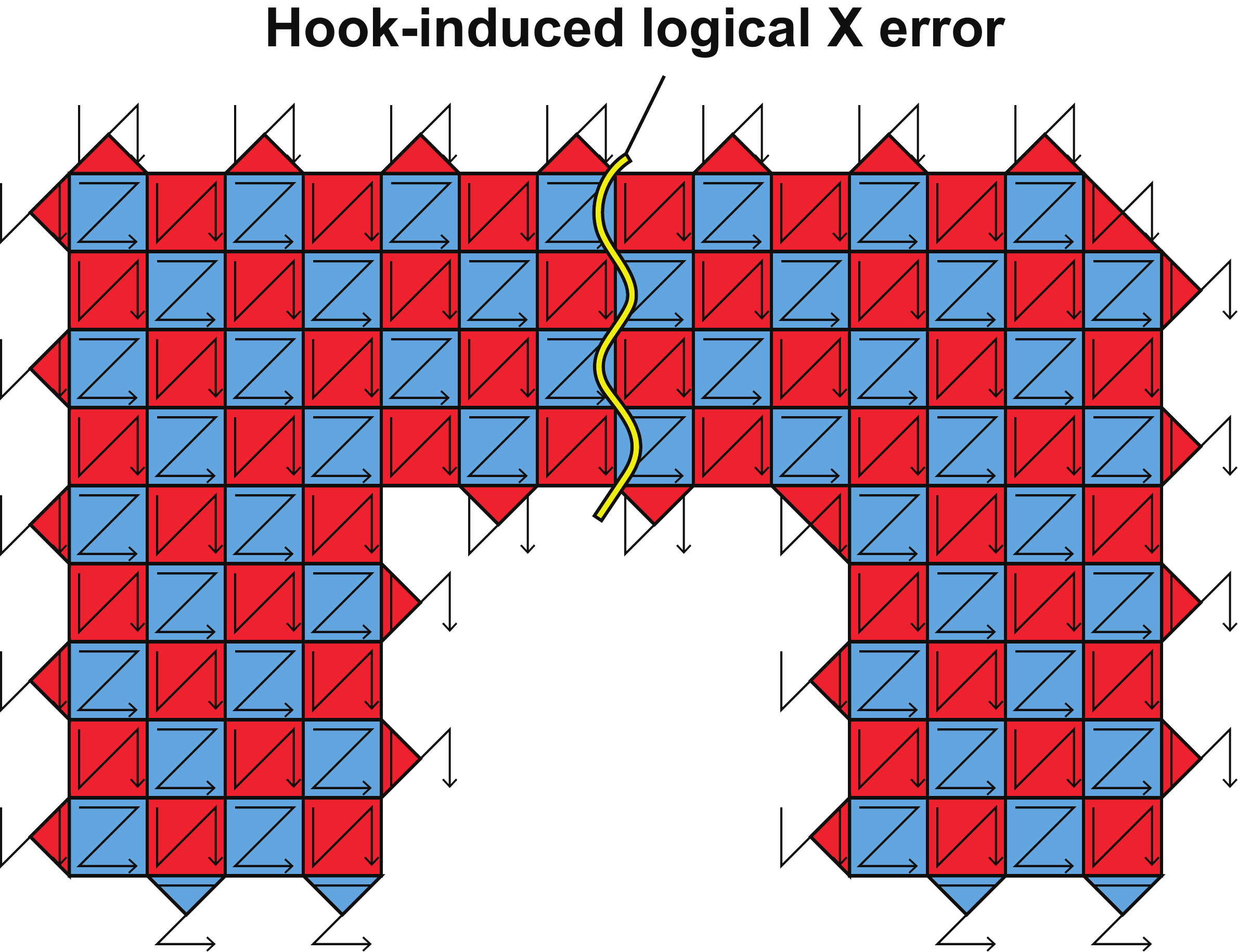}
\caption{An example of the hook-induced logical X error from top to bottom in the logical Pauli-XX measurement depicted by the yellow line.}
\label{fig:hook_prone_logical_x}
\end{figure}

As reviewed in the last section, it is a vital but non-trivial task to provide CNOT ordering that can perform syndrome extraction without any overhead. While there is a preferable CNOT ordering for a memory experiment~\cite{tomita2014low,heim2016optimal}, this cannot be applied to an arbitrary tile layout during lattice surgery, patch rotations, and patch movement. For example, in the layout of Fig.~\ref{fig:XX_measurement}, either of the shortcut paths from left to right and those from top to bottom reduce the fault distance in the X-stabilizer decoding graph. The latter is shown in Fig.~\ref{fig:hook_prone_logical_x}. Thus, the N/Z-shaped ordering and its reflection, rotation, or reversed ones result in the halved fault distance. Providing a systematic construction of CNOT orderings for an arbitrary layout of tiles is strongly demanded for efficient surface-code-based QEC.

More concretely, we aim to design a method for generating a CNOT ordering that satisfies all of the following conditions:
\begin{itemize}
\item CNOT ordering can be determined with a simple rule from the tile layout, i.e., not relying on the expensive methods such as SAT or heuristic search.
\item Fault-distance is equal to the code distance in any regular tile layout. 
\item The length of a syndrome-extraction round does not change. Otherwise, an effective physical error rate per cycle increases, resulting in a degraded reduction rate of logical error probabilities.
\end{itemize}
As explained in the next subsection, to the best of our knowledge, no existing method satisfies all the above requirements.

Note that this paper focuses on layouts in which stabilizers are regularly tiled; here, \textit{regular} means that the surface code consists only of square, tile-like stabilizers, with appropriate two-qubit stabilizers at the boundaries. In this paper, the term \textit{layout} implicitly refers to such a regular tile layout unless otherwise stated. In practice, rectangular stabilizers and stabilizers acting on more than four data qubits can arise \cite{litinski2018lattice}. Supporting such irregular layouts is important but beyond the scope of this paper. Extending the proposed methods to these situations is left as future work.

\subsection{Existing work}
There have been several studies to propose a method to construct distance-preserving syndrome-extraction circuits for an arbitrary tile layout. This section reviews the existing work and shows that there is no method that satisfies the requirements.

Refs.~\cite{gidney2025alternating, bluvstein2025architectural,haug2025lattice} propose an alternating technique. In the methods in Refs.~\cite{gidney2025alternating,haug2025lattice}, the CNOT ordering is N/Z shaped in the even cycles, and its order is reversed in the odd cycles. While this approach achieves a CNOT-depth of four, its fault distance is reduced to $d-1$. In Ref.~\cite{bluvstein2025architectural}, a similar method is used, but the ordering alternates between a Z-shaped ordering in one round and an N-shaped ordering in the next. Its fault distance is also reduced to $d-1$.

In Ref.~\cite{litinski2018lattice}, N/Z shapes are flexibly assigned to prevent shortcut paths induced by hook errors that would otherwise reduce the fault distance. However, this technique increases the depth to avoid multiple CNOT gates simultaneously acting on the same data qubit.

\begin{figure}[t]
    \centering
    \includegraphics[width=0.46\textwidth]{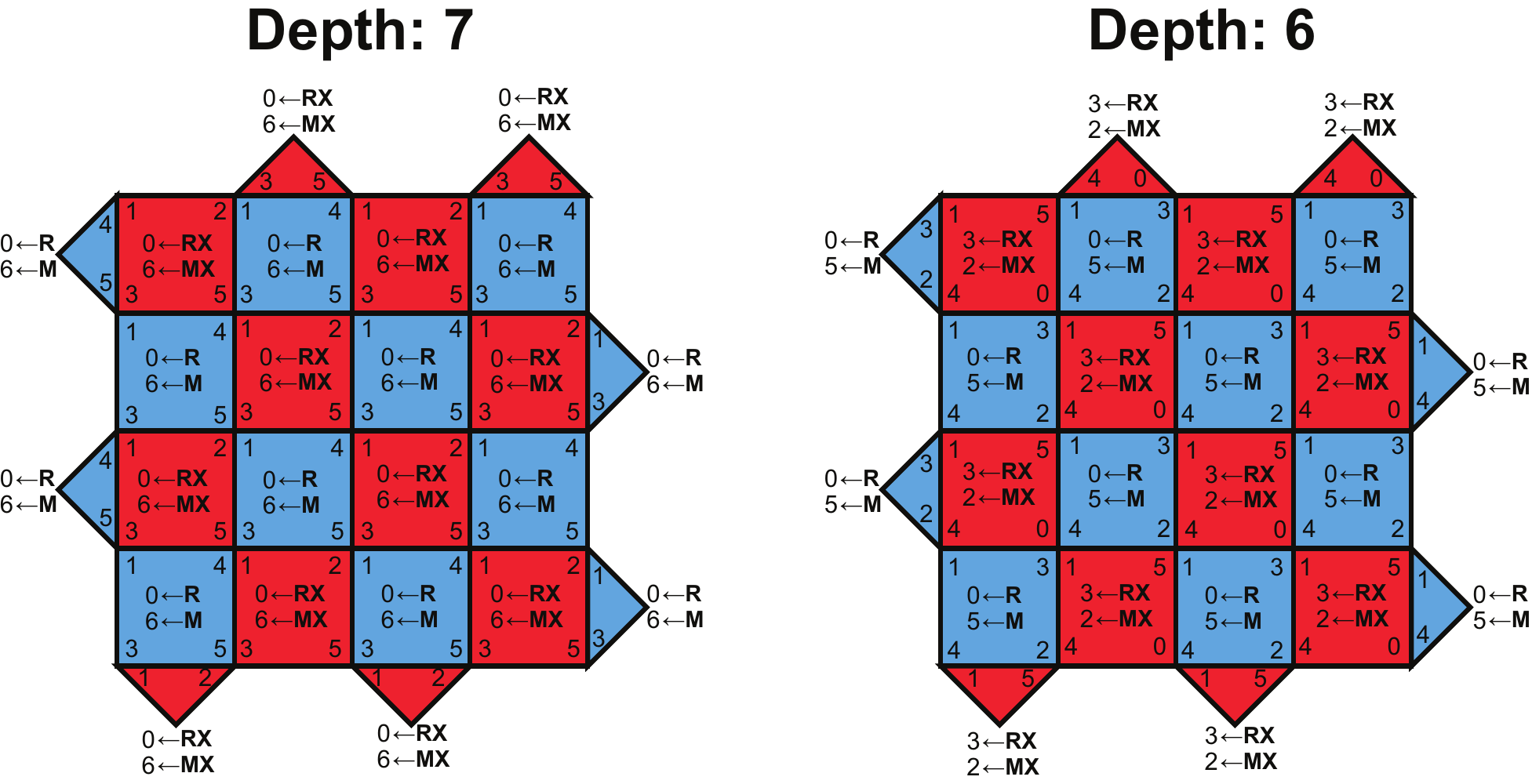}
    \caption{An example of the CNOT ordering in the syndrome extraction proposed in Ref.~\cite{kishony2026surface}. The numbers in the tiles indicate their ordering; the left panel shows the N/Z ordering with an additional depth, and the right panel shows the diagonal ordering with three additional depths. The depths in the figure include the reset and measurement operations, and their CNOT depths (depths excluding reset and measurement) are five and four, respectively.}
    \label{fig:surface_order_kishony}
\end{figure}

Very recently, Ref.~\cite{kishony2026surface} introduced two approaches to avoid this problem independently of our work. One approach is to avoid hook errors by adding an idling depth for N/Z-shaped syndrome measurements, as shown in the left panel of Fig.~\ref{fig:surface_order_kishony}. Here, red tiles are idling at depth four, and blue tiles are at depth two. 

The other approach by Ref.~\cite{kishony2026surface} is shown in the right panel of Fig.~\ref{fig:surface_order_kishony}. It utilizes diagonal ordering to avoid hook errors. While this enables CNOT-depth four, this implementation demands parallel execution of CNOT gates and measurements, which will impose another burden on hardware implementation. For example, if we assume the implementation of superconducting qubits, it becomes difficult to decouple the effects of frequency shifts from readout resonators. Also, in most qubit devices, readout and reset will take a longer time than the CNOT gates~\cite{google2023suppressing,google2025quantum}. Thus, this will increase the duration of syndrome-extraction circuits. If we prohibit the simultaneous execution of measurements and CNOT gates, the CNOT depth increases to seven.

Ref.~\cite{mcewen2023relaxing} proposed another type of syndrome-extraction circuits. While this method does not target mitigating hook-error effects, they propose syndrome-extraction circuits with a full fault distance by letting Z and X tiles move toward the Z and X boundaries, respectively. This enables us to implement surface codes with qubits on hexagonal grids instead of square grids. However, its implementation is dedicated to standard memory layouts and assumes that the Z and X tiles move toward Z and X boundaries, respectively. Thus, we cannot apply this ordering to an arbitrary layout such as Fig.~\ref{fig:XX_measurement}, where the Z and X tiles can move toward X and Z boundaries, respectively. As explained in the following section, our approach is to extend the circuit shown in Fig.~8 of Ref.~\cite{mcewen2023relaxing} to satisfy the requirements.

There are several other proposals that convert a problem of finding a syndrome-extraction circuit with desirable properties into well-known problems and automatically solve them with heuristic solvers~\cite{peham2025automated,liu2026alphasyndrome,viszlai2026prophunt}. While they can be applied to any QEC code, calculating code distances for a given syndrome-extraction circuit is NP-hard, so the runtime for general large layouts is expected to be prohibitively long, limiting the applicable scale. Therefore, there is no systematic design of syndrome-extraction circuits that can achieve the desirable properties of surface codes, and methods for finding them are strongly demanded to straightforwardly demonstrate high-performance QEC experiments.

\section{ZX Interleaving Syndrome Extraction}
\label{sec:zx_interleaving}

Here, we propose a design of a syndrome-extraction circuit satisfying all the requirements explained in the last section, named {\it ZX Interleaving Syndrome Extraction}. This section presents a new method for avoiding hook errors in any layout of surface codes with CNOT depth four. During the design phase of this work, the methods are constructed by using both Crumble and Stim~\cite{gidney_crumble,gidney2021stim}.

\begin{figure*}[htb]
    \centering
    \includegraphics[width=0.9\textwidth]{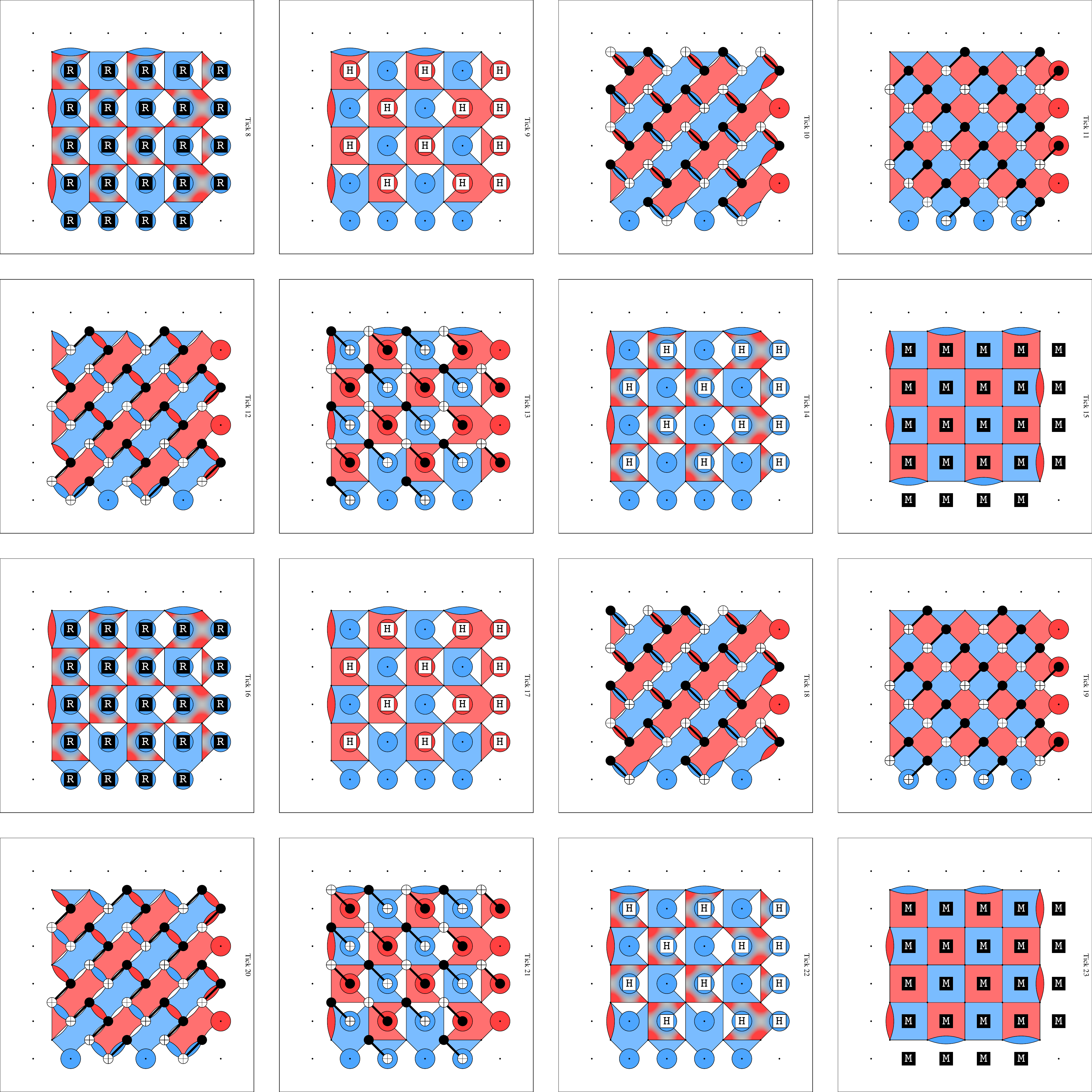}
    \caption{The ZX interleaving syndrome extraction circuit where the Z tiles move toward the Z boundary, and the X tiles move toward the X boundary. \href{https://algassert.com/crumble\#circuit=Q(0,0)0;Q(0,1)1;Q(0,2)2;Q(0,3)3;Q(0,4)4;Q(0.5,0.5)5;Q(0.5,1.5)6;Q(0.5,2.5)7;Q(0.5,3.5)8;Q(0.5,4.5)9;Q(1,0)10;Q(1,1)11;Q(1,2)12;Q(1,3)13;Q(1,4)14;Q(1.5,0.5)15;Q(1.5,1.5)16;Q(1.5,2.5)17;Q(1.5,3.5)18;Q(1.5,4.5)19;Q(2,0)20;Q(2,1)21;Q(2,2)22;Q(2,3)23;Q(2,4)24;Q(2.5,0.5)25;Q(2.5,1.5)26;Q(2.5,2.5)27;Q(2.5,3.5)28;Q(2.5,4.5)29;Q(3,0)30;Q(3,1)31;Q(3,2)32;Q(3,3)33;Q(3,4)34;Q(3.5,0.5)35;Q(3.5,1.5)36;Q(3.5,2.5)37;Q(3.5,3.5)38;Q(3.5,4.5)39;Q(4,0)40;Q(4,1)41;Q(4,2)42;Q(4,3)43;Q(4,4)44;Q(4.5,0.5)45;Q(4.5,1.5)46;Q(4.5,2.5)47;Q(4.5,3.5)48;POLYGON(0,0,1,0.25)11_21_22_12;POLYGON(0,0,1,0.25)22_32_33_23;POLYGON(0,0,1,0.25)13_23_24_14;POLYGON(0,0,1,0.25)33_43_44_34;POLYGON(0,0,1,0.25)31_41_42_32;POLYGON(0,0,1,0.25)20_30_31_21;POLYGON(0,0,1,0.25)0_10_11_1;POLYGON(0,0,1,0.25)2_12_13_3;POLYGON(0,0,1,0.25)20_10;POLYGON(0,0,1,0.25)40_30;POLYGON(0,0,1,0.25)14_4;POLYGON(0,0,1,0.25)34_24;POLYGON(1,0,0,0.25)12_22_23_13;POLYGON(1,0,0,0.25)21_31_32_22;POLYGON(1,0,0,0.25)23_33_34_24;POLYGON(1,0,0,0.25)32_42_43_33;POLYGON(1,0,0,0.25)10_20_21_11;POLYGON(1,0,0,0.25)30_40_41_31;POLYGON(1,0,0,0.25)1_11_12_2;POLYGON(1,0,0,0.25)3_13_14_4;POLYGON(1,0,0,0.25)41_42;POLYGON(1,0,0,0.25)43_44;POLYGON(1,0,0,0.25)0_1;POLYGON(1,0,0,0.25)2_3;TICK;R_5_6_7_8_9_15_16_17_18_19_25_26_27_28_29_35_36_37_38_39_45_46_47_48;MARKZ(0)27;MARKZ(1)17;TICK;H_15_6_26_35_37_28_17_8_48_47_46_45;TICK;CX_8_3_2_7_13_18_24_29_4_9_28_23_17_12_6_1_0_5_11_16_22_27_33_38_48_43_37_32_26_21_15_10_20_25_31_36_46_41_35_30;TICK;CX_10_5_15_11_21_16_26_22_6_2_12_7_17_13_8_4_30_25_35_31_41_36_32_27_23_18_14_9_46_42_37_33_28_24_43_38_34_29_48_44;TICK;CX_20_15_11_6_25_21_16_12_7_3_31_26_40_35_22_17_13_8_45_41_36_32_27_23_18_14_42_37_33_28_24_19_47_43_38_34_44_39_5_1;TICK;CX_5_0_38_33_27_22_16_11_1_6_12_17_23_28_34_39_18_13_7_2_3_8_14_19_10_15_21_26_32_37_36_31_25_20_47_42_30_35_45_40;TICK;H_5_25_16_7_18_27_38_36_45_46_47_48;TICK;M_5_6_7_8_9_15_16_17_18_19_25_26_27_28_29_35_36_37_38_39_45_46_47_48;MARKZ(0)26;MARKZ(1)7;TICK;R_5_6_7_8_9_15_16_17_18_19_25_26_27_28_29_35_36_37_38_39_45_46_47_48;MARKZ(0)26;MARKZ(1)7;MARKZ(2)28;MARKZ(3)27;TICK;H_5_25_16_7_18_27_38_36_45_46_47_48;TICK;CX_38_33_27_22_16_11_5_0_10_15_21_26_32_37_47_42_36_31_25_20_30_35_45_40_1_6_12_17_23_28_34_39_18_13_7_2_3_8_14_19;TICK;CX_16_12_7_3_25_21_31_26_40_35_22_17_13_8_45_41_36_32_18_14_27_23_42_37_33_28_24_19_47_43_38_34_44_39_20_15_11_6_5_1;TICK;CX_26_22_17_13_8_4_35_31_41_36_32_27_23_18_14_9_46_42_37_33_28_24_43_38_34_29_48_44_30_25_21_16_12_7_6_2_15_11_10_5;TICK;CX_0_5_11_16_22_27_33_38_48_43_37_32_26_21_15_10_20_25_31_36_46_41_35_30_28_23_17_12_6_1_2_7_13_18_24_29_8_3_4_9;TICK;H_15_6_26_35_37_28_17_8_48_47_46_45;TICK;M_5_6_7_8_9_15_16_17_18_19_25_26_27_28_29_35_36_37_38_39_45_46_47_48;MARKZ(0)27;MARKZ(1)17;MARKZ(2)27;MARKZ(3)17;TICK;R_5_6_7_8_9_15_16_17_18_19_25_26_27_28_29_35_36_37_38_39_45_46_47_48;MARKZ(0)27;MARKZ(1)17;MARKZ(2)27;MARKZ(3)17;TICK;H_15_6_26_35_37_28_17_8_48_47_46_45;TICK;CX_8_3_2_7_13_18_24_29_4_9_28_23_17_12_6_1_0_5_11_16_22_27_33_38_48_43_37_32_26_21_15_10_20_25_31_36_46_41_35_30;TICK;CX_10_5_15_11_21_16_26_22_6_2_12_7_17_13_8_4_30_25_35_31_41_36_32_27_23_18_14_9_46_42_37_33_28_24_43_38_34_29_48_44;TICK;CX_20_15_11_6_25_21_16_12_7_3_31_26_40_35_22_17_13_8_45_41_36_32_27_23_18_14_42_37_33_28_24_19_47_43_38_34_44_39_5_1;TICK;CX_5_0_38_33_27_22_16_11_1_6_12_17_23_28_34_39_18_13_7_2_3_8_14_19_10_15_21_26_32_37_36_31_25_20_47_42_30_35_45_40;TICK;H_5_25_16_7_18_27_38_36_45_46_47_48;TICK;M_5_6_7_8_9_15_16_17_18_19_25_26_27_28_29_35_36_37_38_39_45_46_47_48;MARKZ(0)26;MARKZ(1)7;MARKZ(2)28;MARKZ(3)27;TICK;R_5_6_7_8_9_15_16_17_18_19_25_26_27_28_29_35_36_37_38_39_45_46_47_48;MARKZ(0)26;MARKZ(1)7;TICK;H_5_25_16_7_18_27_38_36_45_46_47_48;TICK;CX_38_33_27_22_16_11_5_0_10_15_21_26_32_37_47_42_36_31_25_20_30_35_45_40_1_6_12_17_23_28_34_39_18_13_7_2_3_8_14_19;TICK;CX_16_12_7_3_25_21_31_26_40_35_22_17_13_8_45_41_36_32_18_14_27_23_42_37_33_28_24_19_47_43_38_34_44_39_20_15_11_6_5_1;TICK;CX_26_22_17_13_8_4_35_31_41_36_32_27_23_18_14_9_46_42_37_33_28_24_43_38_34_29_48_44_30_25_21_16_12_7_6_2_15_11_10_5;TICK;CX_0_5_11_16_22_27_33_38_48_43_37_32_26_21_15_10_20_25_31_36_46_41_35_30_28_23_17_12_6_1_2_7_13_18_24_29_8_3_4_9;TICK;H_15_6_26_35_37_28_17_8_48_47_46_45;TICK;M_5_6_7_8_9_15_16_17_18_19_25_26_27_28_29_35_36_37_38_39_45_46_47_48;MARKZ(0)27;MARKZ(1)17}{Click here to open this circuit in Crumble.}}
    \label{fig:Z2Z_X2X}
\end{figure*}

Our idea is based on the decoding-graph analysis of the syndrome-extraction circuits proposed in Ref.~\cite{mcewen2023relaxing} for removing the effect of hook errors without increasing the CNOT depth. Let us first review the syndrome extraction circuit proposed in Ref.~\cite{mcewen2023relaxing} in the view of decoding graphs. Figure~\ref{fig:Z2Z_X2X} shows a detector diagram of the syndrome extraction circuits in Ref.~\cite{mcewen2023relaxing}. In this scheme, the Z and X tiles interleave in every round of syndrome extraction: the Z tiles shift upward toward the Z boundary, while the X tiles shift leftward toward the X boundary. The CNOT ordering is neither N-shaped nor Z-shaped; it is a new ordering for the surface code that achieves minimum depth for the syndrome extraction. In particular, the last two CNOT gates in the ordering prevent hook errors from contaminating the logical operators, thereby preserving the fault distance.

\begin{figure}[tb]
    \centering
    \includegraphics[width=0.48\textwidth]{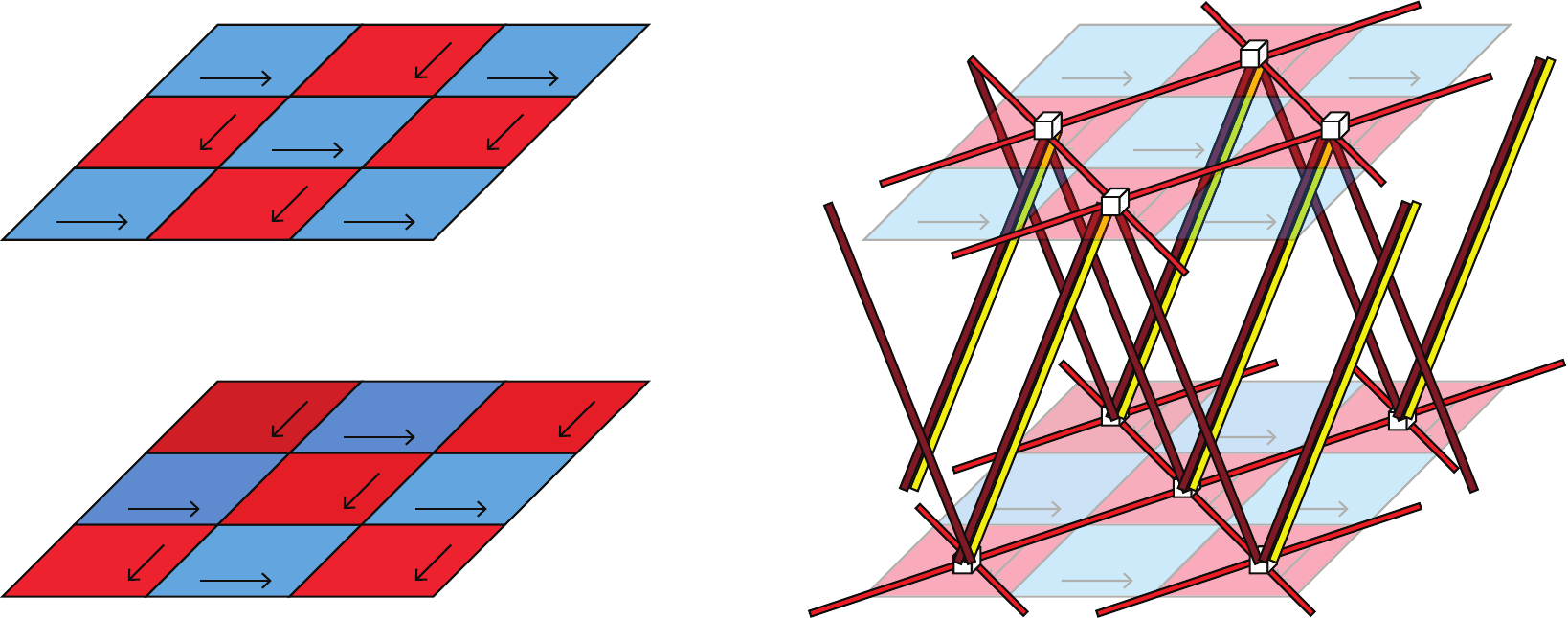}
    \caption{Edges corresponding to hook errors in the X-stabilizer decoding graph of the ZX interleaving syndrome extraction circuit. The light red edges represent the errors on the data qubits, the dark red edges represent the errors on the measurement qubits, and the yellow edges represent hook-error events arising from the syndrome extraction of the Z stabilizers. The edges corresponding to the measurement errors overlap with those of the hook errors. Some error events are omitted for clarity. The arrows represent the ordering of the last two CNOT gates.}
    \label{fig:hook_errors_in_dec_ZX_interleaved}
\end{figure}

A closer look reveals an additional insight. Figure~\ref{fig:hook_errors_in_dec_ZX_interleaved} shows a focused X-stabilizer decoding graph. In the decoding graph, the edges corresponding to hook errors are shortened. This occurs because the tiles move parallel to the hook errors but in the opposite direction. As a result, hook errors in this circuit become harmless regardless of their direction in the surface code; effectively, they no longer exist.

\begin{figure}[tb]
    \centering
    \includegraphics[width=0.45\textwidth]{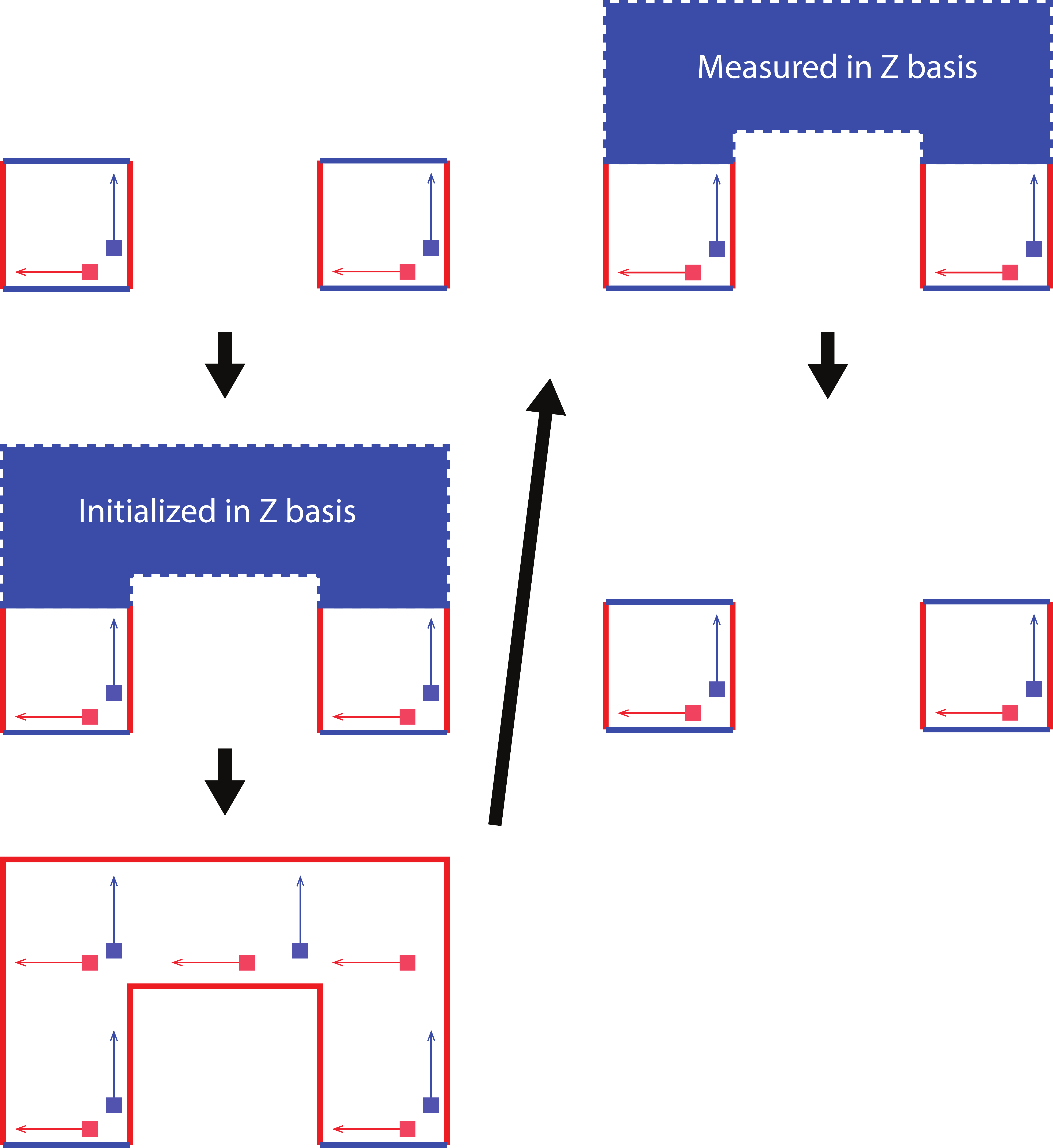}
    \caption{Tile movement during lattice-surgery operations. Red and blue tiles correspond to Z and X tiles, respectively.}
    \label{fig:detector_diagram_XX_meas_movement}
\end{figure}

\begin{figure*}[t]
    \centering
    \includegraphics[width=0.9\textwidth]{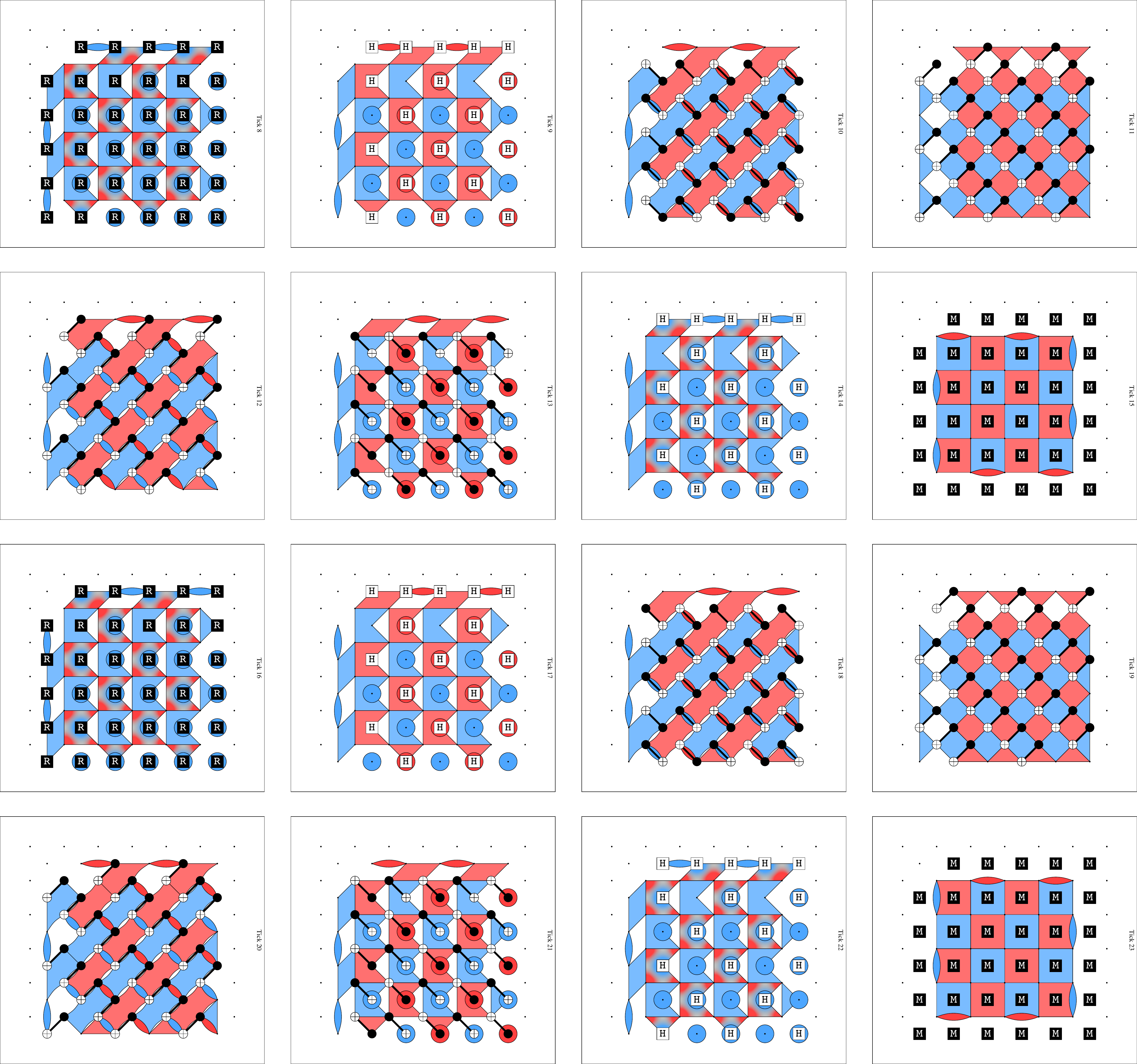}
    \caption{The ZX interleaving syndrome extraction circuit where the Z tiles move toward the X boundary, and the X tiles move toward the Z boundary. \href{https://algassert.com/crumble\#circuit=Q(0,1)0;Q(0,2)1;Q(0,3)2;Q(0,4)3;Q(0,5)4;Q(0.5,0.5)5;Q(0.5,1.5)6;Q(0.5,2.5)7;Q(0.5,3.5)8;Q(0.5,4.5)9;Q(1,0)10;Q(1,1)11;Q(1,2)12;Q(1,3)13;Q(1,4)14;Q(1,5)15;Q(1.5,0.5)16;Q(1.5,1.5)17;Q(1.5,2.5)18;Q(1.5,3.5)19;Q(1.5,4.5)20;Q(2,0)21;Q(2,1)22;Q(2,2)23;Q(2,3)24;Q(2,4)25;Q(2,5)26;Q(2.5,0.5)27;Q(2.5,1.5)28;Q(2.5,2.5)29;Q(2.5,3.5)30;Q(2.5,4.5)31;Q(3,0)32;Q(3,1)33;Q(3,2)34;Q(3,3)35;Q(3,4)36;Q(3,5)37;Q(3.5,0.5)38;Q(3.5,1.5)39;Q(3.5,2.5)40;Q(3.5,3.5)41;Q(3.5,4.5)42;Q(4,0)43;Q(4,1)44;Q(4,2)45;Q(4,3)46;Q(4,4)47;Q(4,5)48;Q(4.5,0.5)49;Q(4.5,1.5)50;Q(4.5,2.5)51;Q(4.5,3.5)52;Q(4.5,4.5)53;Q(5,0)54;Q(5,1)55;Q(5,2)56;Q(5,3)57;Q(5,4)58;Q(5,5)59;POLYGON(0,0,1,0.25)5_16_17_6;POLYGON(0,0,1,0.25)17_28_29_18;POLYGON(0,0,1,0.25)29_40_41_30;POLYGON(0,0,1,0.25)41_52_53_42;POLYGON(0,0,1,0.25)27_38_39_28;POLYGON(0,0,1,0.25)39_50_51_40;POLYGON(0,0,1,0.25)19_30_31_20;POLYGON(0,0,1,0.25)7_18_19_8;POLYGON(0,0,1,0.25)6_7;POLYGON(0,0,1,0.25)8_9;POLYGON(0,0,1,0.25)49_50;POLYGON(0,0,1,0.25)51_52;POLYGON(1,0,0,0.25)40_51_52_41;POLYGON(1,0,0,0.25)28_39_40_29;POLYGON(1,0,0,0.25)16_27_28_17;POLYGON(1,0,0,0.25)38_49_50_39;POLYGON(1,0,0,0.25)6_17_18_7;POLYGON(1,0,0,0.25)18_29_30_19;POLYGON(1,0,0,0.25)30_41_42_31;POLYGON(1,0,0,0.25)8_19_20_9;POLYGON(1,0,0,0.25)16_5;POLYGON(1,0,0,0.25)38_27;POLYGON(1,0,0,0.25)31_20;POLYGON(1,0,0,0.25)53_42;TICK;R_11_12_13_14_22_23_24_25_33_34_35_36_44_45_46_47_10_21_32_43_54_55_56_57_58_59_4_15_26_37_48_0_1_2_3;MARKZ(0)35;MARKZ(1)24;MARKZ(4)2_3;MARKZ(6)21_32;TICK;H_36_46_24_14_12_22_34_44_56_58_10_21_32_43_54_48_26;TICK;CX_5_11_17_23_29_35_41_47_34_28_22_16_46_40_58_52_27_33_39_45_51_57_56_50_44_38_49_55_53_59_48_42_36_30_24_18_12_6_7_13_19_25_31_37_26_20_14_8_9_15;TICK;CX_16_11_22_17_12_7_10_5_32_27_38_33_28_23_18_13_8_3_54_49_44_39_34_29_24_19_14_9_50_45_40_35_30_25_20_15_56_51_46_41_36_31_52_47_42_37_58_53_6_1;TICK;CX_11_6_17_12_23_18_29_24_35_30_41_36_47_42_53_48_5_0_21_16_27_22_33_28_39_34_45_40_51_46_57_52_55_50_49_44_43_38_7_2_13_8_19_14_25_20_31_26_9_4;TICK;CX_11_5_23_17_47_41_35_29_59_53_16_22_28_34_40_46_52_58_57_51_45_39_33_27_38_44_50_56_55_49_6_12_18_24_30_36_42_48_37_31_25_19_13_7_8_14_20_26_15_9;TICK;H_11_23_35_47_25_13_45_33_10_21_32_43_54_57_55_37_15;TICK;M_0_1_2_3_4_15_26_37_48_59_10_11_12_13_14_21_22_23_24_25_32_33_34_35_36_43_44_45_46_47_54_55_56_57_58;MARKZ(0)34;MARKZ(1)13;MARKZ(4)3_2;MARKZ(6)21_32;TICK;R_0_1_2_3_4_10_11_12_13_14_15_21_22_23_24_25_26_32_33_34_35_36_37_43_44_45_46_47_48_54_55_56_57_58_59;MARKZ(0)34;MARKZ(1)13;MARKZ(2)36;MARKZ(3)35;MARKZ(4)2_3;MARKZ(5)1_2;MARKZ(6)21_32;MARKZ(7)32_43;TICK;H_11_23_35_47_25_13_45_33_10_21_32_43_54_57_55_37_15;TICK;CX_6_12_18_24_30_36_42_48_37_31_25_19_13_7_8_14_20_26_15_9_59_53_47_41_35_29_23_17_11_5_16_22_28_34_40_46_52_58_57_51_45_39_33_27_38_44_50_56_55_49;TICK;CX_27_22_33_28_39_34_45_40_51_46_41_36_31_26_25_20_35_30_29_24_19_14_9_4_23_18_13_8_17_12_7_2_21_16_11_6_5_0_49_44_43_38_55_50_57_52_47_42_53_48;TICK;CX_28_23_18_13_8_3_38_33_54_49_44_39_34_29_24_19_14_9_50_45_40_35_30_25_20_15_56_51_46_41_36_31_52_47_42_37_58_53_32_27_22_17_12_7_16_11_6_1_10_5;TICK;CX_5_11_17_23_29_35_41_47_53_59_48_42_36_30_24_18_12_6_7_13_19_25_31_37_26_20_14_8_9_15_58_52_46_40_34_28_22_16_27_33_39_45_51_57_56_50_44_38_49_55;TICK;H_36_46_24_14_12_22_34_44_56_58_10_21_32_43_54_48_26;TICK;M_0_1_2_3_4_15_26_37_48_59_10_11_12_13_14_21_22_23_24_25_32_33_34_35_36_43_44_45_46_47_54_55_56_57_58;MARKZ(0)35;MARKZ(1)24;MARKZ(2)35;MARKZ(3)24;MARKZ(4)2_3;MARKZ(5)1_2;MARKZ(6)21_32;MARKZ(7)32_43;TICK;R_0_1_2_3_4_15_26_37_48_59_10_11_12_13_14_21_22_23_24_25_32_33_34_35_36_43_44_45_46_47_54_55_56_57_58;MARKZ(0)35;MARKZ(1)24;MARKZ(2)35;MARKZ(3)24;MARKZ(4)2_3;MARKZ(5)1_2;MARKZ(6)21_32;MARKZ(7)32_43;TICK;H_36_46_24_14_12_22_34_44_56_58_10_21_32_43_54_48_26;TICK;CX_5_11_17_23_29_35_41_47_34_28_22_16_46_40_58_52_27_33_39_45_51_57_56_50_44_38_49_55_53_59_48_42_36_30_24_18_12_6_7_13_19_25_31_37_26_20_14_8_9_15;TICK;CX_16_11_22_17_12_7_10_5_32_27_38_33_28_23_18_13_8_3_54_49_44_39_34_29_24_19_14_9_50_45_40_35_30_25_20_15_56_51_46_41_36_31_52_47_42_37_58_53_6_1;TICK;CX_11_6_17_12_23_18_29_24_35_30_41_36_47_42_53_48_5_0_21_16_27_22_33_28_39_34_45_40_51_46_57_52_55_50_49_44_43_38_7_2_13_8_19_14_25_20_31_26_9_4;TICK;CX_11_5_23_17_47_41_35_29_59_53_16_22_28_34_40_46_52_58_57_51_45_39_33_27_38_44_50_56_55_49_6_12_18_24_30_36_42_48_37_31_25_19_13_7_8_14_20_26_15_9;TICK;H_11_23_35_47_25_13_45_33_10_21_32_43_54_57_55_37_15;TICK;M_0_1_2_3_4_15_26_37_48_59_10_11_12_13_14_21_22_23_24_25_32_33_34_35_36_43_44_45_46_47_54_55_56_57_58;MARKZ(0)34;MARKZ(1)13;MARKZ(2)36;MARKZ(3)35;MARKZ(4)2_3;MARKZ(5)1_2;MARKZ(6)21_32;MARKZ(7)32_43;TICK;R_0_1_2_3_4_10_11_12_13_14_15_21_22_23_24_25_26_32_33_34_35_36_37_43_44_45_46_47_48_54_55_56_57_58_59;MARKZ(0)34;MARKZ(1)13;MARKZ(4)2_3;MARKZ(6)21_32;TICK;H_11_23_35_47_25_13_45_33_10_21_32_43_54_57_55_37_15;TICK;CX_6_12_18_24_30_36_42_48_37_31_25_19_13_7_8_14_20_26_15_9_59_53_47_41_35_29_23_17_11_5_16_22_28_34_40_46_52_58_57_51_45_39_33_27_38_44_50_56_55_49;TICK;CX_27_22_33_28_39_34_45_40_51_46_41_36_31_26_25_20_35_30_29_24_19_14_9_4_23_18_13_8_17_12_7_2_21_16_11_6_5_0_49_44_43_38_55_50_57_52_47_42_53_48;TICK;CX_28_23_18_13_8_3_38_33_54_49_44_39_34_29_24_19_14_9_50_45_40_35_30_25_20_15_56_51_46_41_36_31_52_47_42_37_58_53_32_27_22_17_12_7_16_11_6_1_10_5;TICK;CX_5_11_17_23_29_35_41_47_53_59_48_42_36_30_24_18_12_6_7_13_19_25_31_37_26_20_14_8_9_15_58_52_46_40_34_28_22_16_27_33_39_45_51_57_56_50_44_38_49_55;TICK;H_36_46_24_14_12_22_34_44_56_58_10_21_32_43_54_48_26;TICK;M_0_1_2_3_4_15_26_37_48_59_10_11_12_13_14_21_22_23_24_25_32_33_34_35_36_43_44_45_46_47_54_55_56_57_58;MARKZ(0)35;MARKZ(1)24;MARKZ(4)2_3;MARKZ(6)21_32}{Click here to open this circuit in Crumble.}}
    \label{fig:Z2X_X2Z}
\end{figure*}

Now, we apply this method to the arbitrary layouts of the surface code. However, Z tiles and X tiles do not always move toward the same type of boundaries, which means that there exist cases in which they move toward opposite types of boundaries. For example, when we apply this approach to lattice-surgery layout, Z tiles must move towards both Z and X boundaries, as shown in Fig.~\ref{fig:detector_diagram_XX_meas_movement}. So we must also consider an additional pattern of the ZX interleaving scheme. To this end, we propose a construction for this case at minimum depth, shown in Fig.~\ref{fig:Z2X_X2Z}. This circuit requires more measurement qubits at the boundaries than the previous case, but it preserves the fault distance. 

\begin{figure*}[t]
    \centering
    \includegraphics[width=0.9\textwidth]{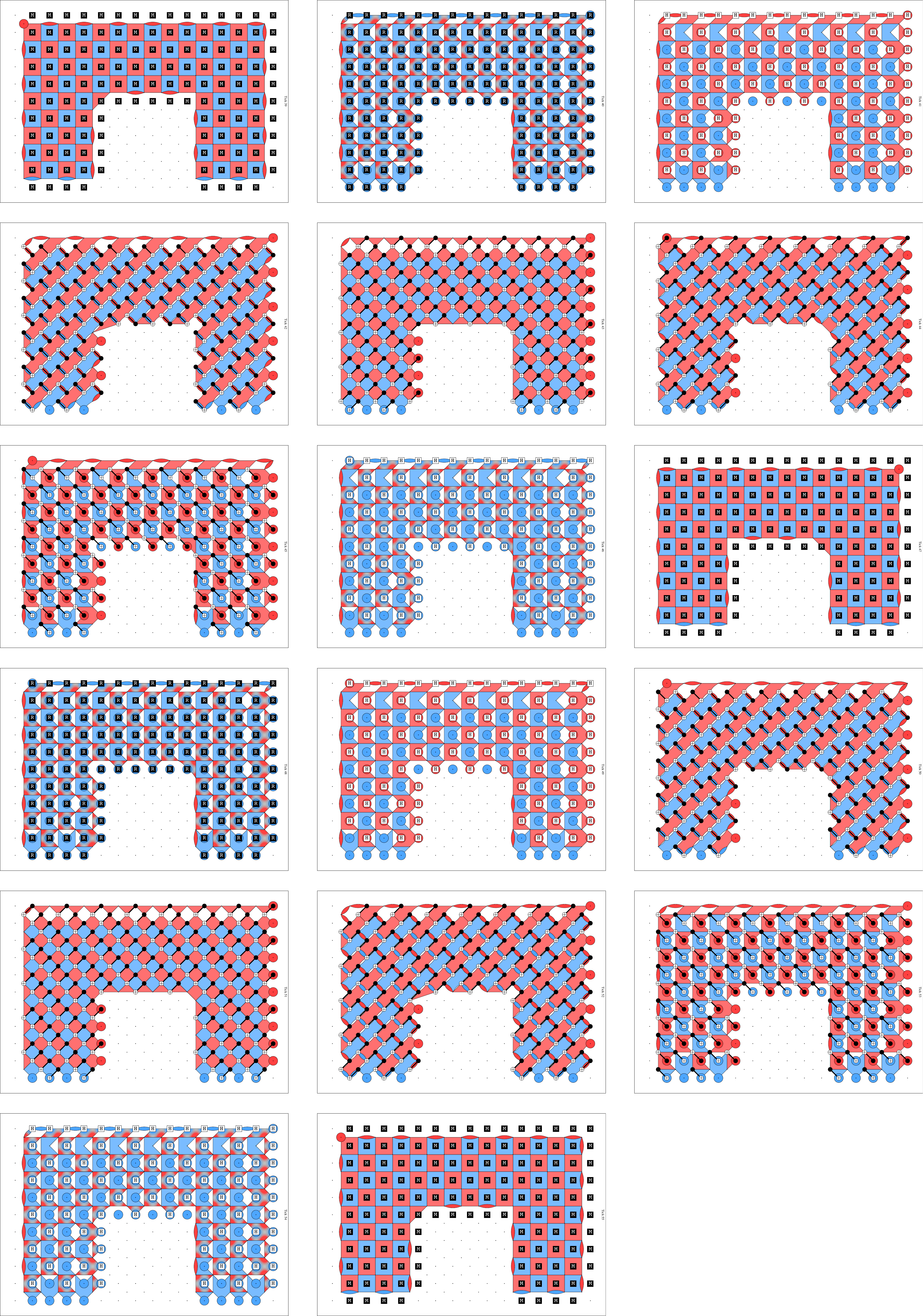}
    \caption{Detector diagram of the ZX interleaving syndrome extraction circuit for the lattice-surgery layout.}
    \label{fig:detector_diagram_XX_meas}
\end{figure*}

By combining these ZX interleaving methods, any layout of the surface code can be realized while preserving the fault distance at the minimum depth. This applies, for example, to the layout shown in Fig.~\ref{fig:XX_measurement}. The detector diagrams of one round of syndrome-extraction circuits are shown in Fig.~\ref{fig:detector_diagram_XX_meas}. Unlike existing methods, the proposed method never requires additional circuit depth, and the CNOT ordering is uniform across the entire surface code, making the construction straightforward for any layout. The proposed method does not require the simultaneous execution of measurements or resets with CNOT gates. Also, this method inherits the hexagonal-grid implementation of surface codes proposed in Ref.~\cite{mcewen2023relaxing}, i.e., each physical qubit is coupled to only three surrounding qubits, which removes unexpected crosstalk and improves their fidelities. Note that although the proposed method uses additional qubits, they are originally unused measurement qubits and are located on the boundaries. Therefore, its logical-qubit density is equal to that of the N/Z ordering.

\section{Numerical Results}
\label{sec:numerical}

\subsection{Benchmark settings}
This section benchmarks the proposed method with the existing methods and evaluates their performance. 
We compared the proposed method with the syndrome-extraction circuits using four layers of CNOT gates, with N/Z ordering \cite{tomita2014low,heim2016optimal} and forward-and-backward alternating ordering~\cite{gidney2025alternating,haug2025lattice}. We implemented these methods in Stim~\cite{gidney2021stim} and evaluated them with a circuit-level noise model, i.e., one- and two-qubit uniform depolarizing noise acts on the target qubits of each one- and two-qubit gate, respectively. We investigate the performance for two types of layouts: memory experiments~(Fig.~\ref{fig:def_3D_diagram}) and lattice surgery for Pauli-XX measurements~(Fig.~\ref{fig:XX_meas}). All circuits are decoded by using the Python library PyMatching~\cite{higgott2022pymatching}, and the simulations are executed in the Python library Sinter \cite{sinter}. Our Stim circuits are available on \href{https://github.com/yugahirai/num_exp_no_more_hooks_in_the_surface_code}{GitHub repository}~\cite{stimcircuit_github}.

\subsection{Memory-experiment layout}
We first evaluate the performance for memory experiments~(Fig.~\ref{fig:def_3D_diagram}) with $d$ rounds of the syndrome extraction. We have evaluated the fault distance of syndrome-extraction circuits with \texttt{shortest\_graphlike\_error} function in Stim~\cite{gidney2021stim}, which provides the upper-bound of fault distances. The results are listed in Table~\ref{tab:surface_code_fault_distance}. N/Z ordering (hook-avoiding) and  N/Z ordering (hook-prone) are syndrome-extraction methods in which the direction of shortcut paths by hook errors is orthogonal to and parallel to the logical operators, respectively. As expected, the fault distance is halved if shortcut paths are parallel to logical operators. As reported in Refs.~\cite{gidney2025alternating,haug2025lattice}, the alternating method loses a fault distance by one. In contrast, the two patterns of the ZX-interleaving method, Z~(X) tile moves to Z~(X) boundary (Z$\to$Z, X$\to$X, Fig.~\ref{fig:Z2Z_X2X}) and Z~(X) tiles move to X~(Z) boundary (Z$\to$X, X$\to$Z, Fig.~\ref{fig:Z2X_X2Z}), achieve the full fault distance for memory-experiment layouts.

As the \texttt{shortest\_graphlike\_error} function provides the upper-bound of fault distance and does not provide the exact one, we verified that our proposal keeps a full fault distance with the heuristic search function \texttt{search\_for\_undetectable\_logical\_errors}. We chose the parameters as follows:
\begin{lstlisting}[basicstyle=\ttfamily\scriptsize]
circuit.search_for_undetectable_logical_errors(
    dont_explore_detection_event_sets_with_size_above=d,
    dont_explore_edges_with_degree_above=5,
    dont_explore_edges_increasing_symptom_degree=False,
    canonicalize_circuit_errors=False,
)
\end{lstlisting}

\begin{figure*}[tb]
    \centering
    \includegraphics[width=1\textwidth]{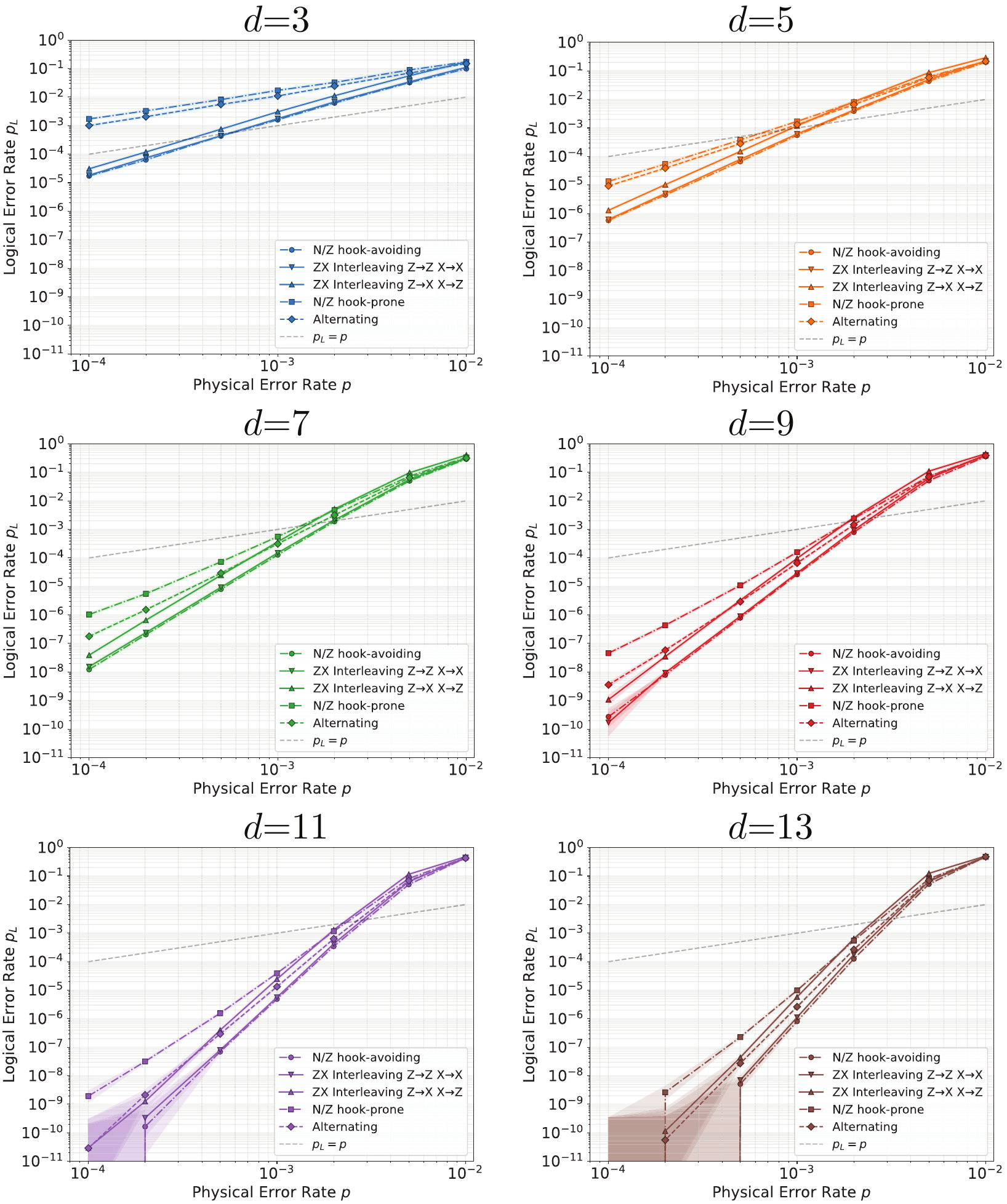}
    \caption{Logical error rates of the memory experiments.}
    \label{fig:comparison_memory} 
\end{figure*}

Next, we evaluate logical error rates for several code distances. The results are shown in Fig.~\ref{fig:comparison_memory}. 
We can see the hook-prone N/Z ordering performs worst, as expected from its fault distance of $d/2$. The hook-avoiding N/Z ordering performs the best owing to its full fault distance of $d$ and simpler circuit design. The ZX interleaving ordering shows a similar scaling to that of the hook-avoiding N/Z ordering, confirming that the fault distance of our proposal indeed achieves $d$.

\begin{table}[t]
    \caption{Fault distance for memory-experiment layout}
    \label{tab:surface_code_fault_distance}
    \begin{tabular}{|c|c|}
        \hline
        Method & Fault distance \\
        \hline
        \hline
        \makecell{\bf N/Z ordering \cite{tomita2014low,heim2016optimal} \\ \bf (hook-avoiding)} & $d$ \\
        \hline
        \makecell{\bf N/Z ordering \\ \bf (hook-prone)} & $d/2$ \\
        \hline
        \bf alternating \cite{gidney2025alternating,haug2025lattice} & $d-1$ \\
        \hline
        \makecell{\bf ZX interleaving (Z$\to$Z, X$\to$X) \cite{mcewen2023relaxing} \\ (Fig.~\ref{fig:Z2Z_X2X})}  & $d$ \\
        \hline
        \makecell{\bf ZX interleaving (Z$\to$X, X$\to$Z) \\ (Proposed, Fig.~\ref{fig:Z2X_X2Z})} & $d$ \\
        \hline
    \end{tabular}
\end{table}

\begin{table}[t]
    \caption{Fault distance for lattice-surgery layout}
    \label{tab:lattice_surgery_fault_distance}
    \resizebox{\columnwidth}{!}{
    \begin{tabular}{|c|c|c|}
        \hline
        Method & Fault distance (X) & Fault distance (Z) \\
        \hline
        \hline
        \bf N/Z ordering \cite{tomita2014low,heim2016optimal} & $d/2$ & $d$ \\
        \hline
        \bf alternating \cite{gidney2025alternating,haug2025lattice} & $d-1$ & $d$ \\
        \hline
        \makecell{\bf ZX interleaving \\ (Proposed)} & $d$ & $d$ \\
        \hline
    \end{tabular}
    }
\end{table}

\begin{figure*}[tb]
    \centering
    \includegraphics[width=1\textwidth]{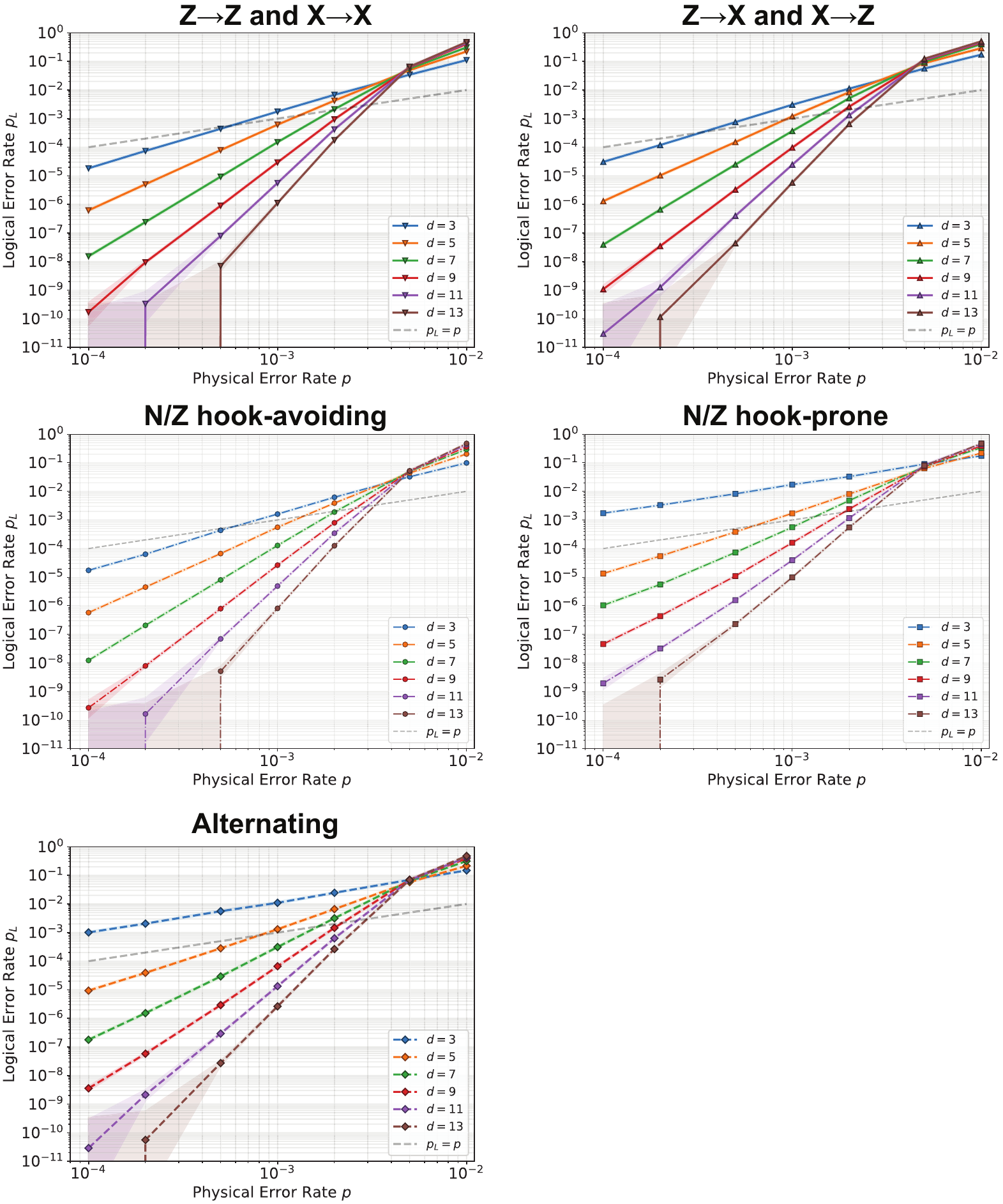}
    \caption{Comparison of the threshold of the memory experiments for all methods.}
    \label{fig:comparison_threshold} 
\end{figure*}

\begin{figure*}[tb]
    \centering
    \includegraphics[width=1\textwidth]{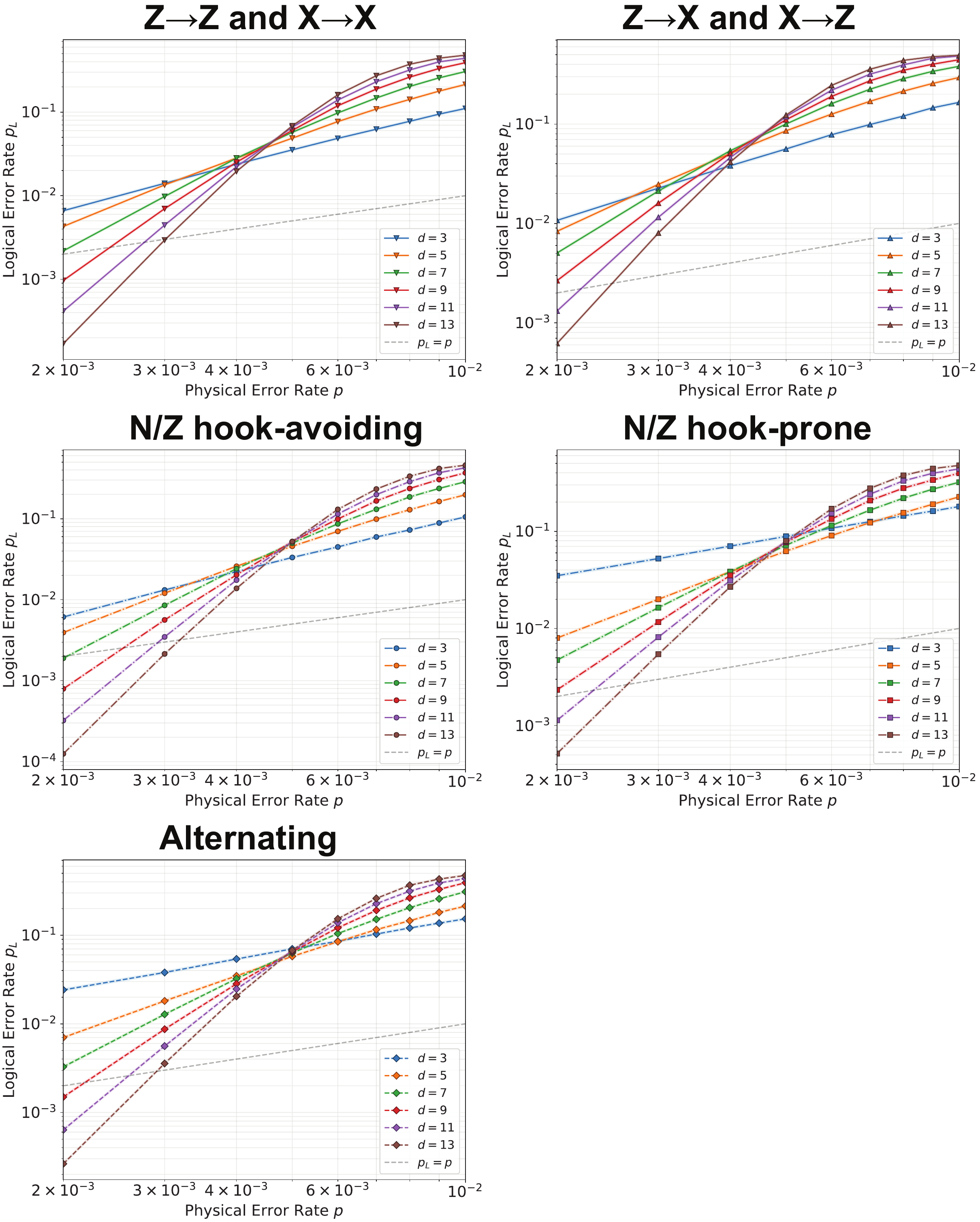}
    \caption{Comparison of the threshold of the memory experiments for all methods focusing on the thresholds.}
    \label{fig:comparison_threshold_close} 
\end{figure*}

When we compare the ZX interleaving ordering and the alternating ordering, for all code distances, the Z$\to$Z and X$\to$X variant achieves the lowest logical error rate. At small code distances or low physical error rates, the Z$\to$X and X$\to$Z variant also outperforms the alternating ordering; however, at large code distances or high physical error rates, the alternating ordering shows the lower logical error rate. We attribute this crossover to the fact that the ZX interleaving ordering introduces additional error events triggered by measurement errors. These events significantly contribute to the logical error rate in the Z$\to$X and X$\to$Z variant because the tiles shift toward the logical operators. No such error events arise in the alternating ordering, and such an error does not significantly contribute to the logical error rates of Z$\to$Z and X$\to$X variant. Consequently, the threshold of the Z$\to$X and X$\to$Z variant may be lower than those of the alternating ordering and the Z$\to$Z and X$\to$X variant. To confirm this, we also plot logical error rates for each method to evaluate the threshold values as shown in Fig.~\ref{fig:comparison_threshold} and~\ref{fig:comparison_threshold_close}. We can see that the threshold of the Z$\to$X and X$\to$Z variant is slightly lower than those of the other methods, including the hook-prone N/Z ordering, and this confirms the above discussion. Interestingly, even the threshold of the Z$\to$Z and X$\to$X variant is slightly lower than those of the other methods, excluding the Z$\to$X and X$\to$Z variant. One possible explanation is that error events arising from measurement errors combined with tile movement are less harmful in the low physical error rate regime; however, long error chains involving such events become non-negligible in the high physical error rate regime. These are drawbacks of our implementation, but we expect they can be compensated for by the fidelity improvement from the advantage of hexagonal-grid qubit allocations. It was experimentally reported that the surface code on a hexagonal grid can achieve a lower logical error rate than that on a square grid~\cite{eickbusch2025demonstration}.

\subsection{Lattice-surgery layout}
We evaluated the fault distance of the lattice-surgery layout for logical Pauli-XX measurements shown in Fig.~\ref{fig:XX_measurement} and Fig.~\ref{fig:XX_meas}.
As X and Z errors are asymmetric for this experiment, we evaluated fault distance for X and Z errors independently. The results are shown in Table~\ref{tab:lattice_surgery_fault_distance}. All methods achieve full fault distance for Z errors. On the other hand, for X errors, the N/Z ordering shows halved fault distances. This is because there are horizontal and vertical logical X operators in this setting, and the N/Z ordering cannot eliminate problematic shortcut paths by hook errors. The alternating ordering mitigates this problem, but there is a loss of fault distance by one. In contrast, the proposed method shows the full fault distance for this layout. To the best of our knowledge, this is the first result that shows full fault distance with no overheads on the length of syndrome extraction circuits. We also validated the full fault distance of our proposal with the function \texttt{search\_for\_undetectable\_logical\_errors} using the same settings as the memory-experiment layout.

\begin{figure*}[tb]
    \centering
    \includegraphics[width=0.9\textwidth]{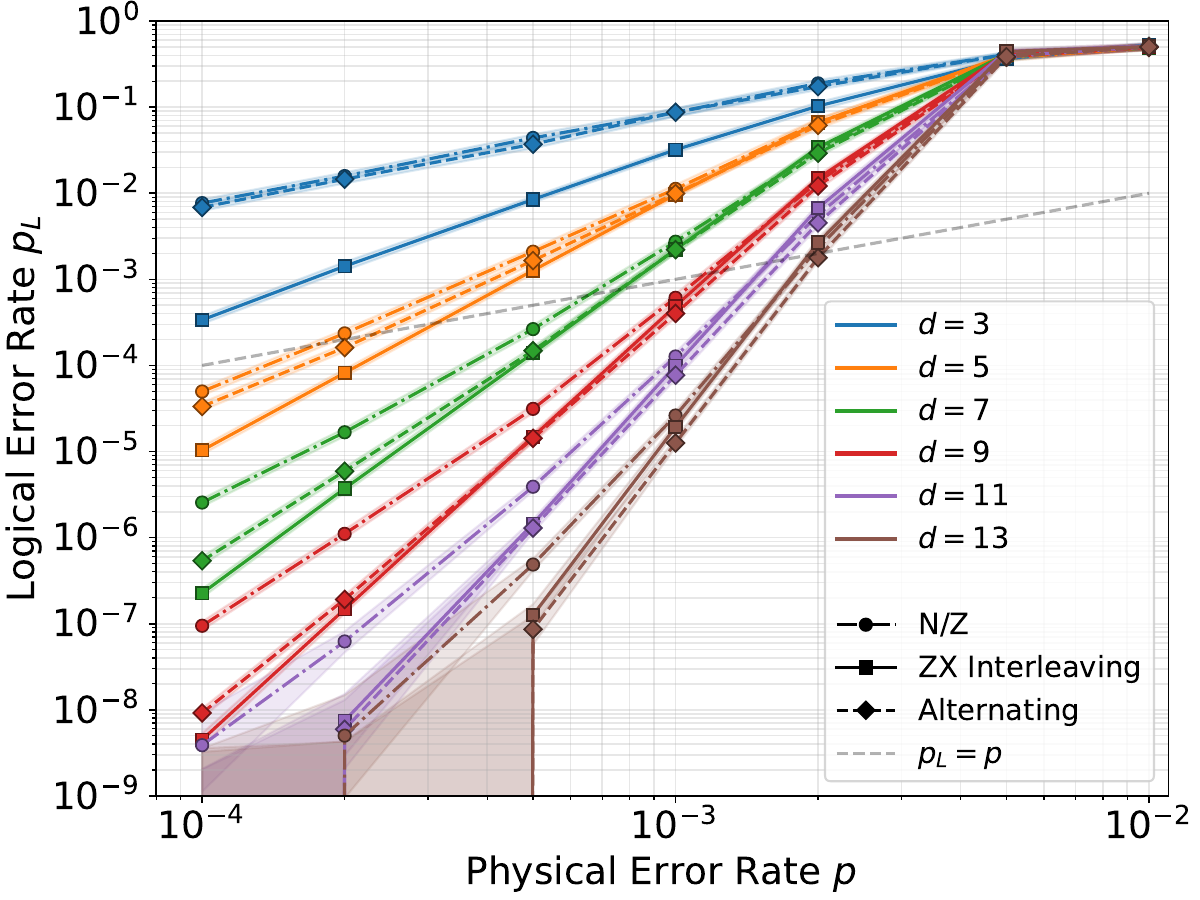}
    \caption{Logical error rates of the lattice-surgery experiments}
    \label{fig:comparison_XX_meas} 
\end{figure*}

We then evaluate the logical error rates against X errors for the lattice-surgery layout. The results are shown in Fig.~\ref{fig:comparison_XX_meas}.
As in the memory experiments, the ZX interleaving ordering achieves a lower logical error rate than the alternating ordering when the physical error rate is sufficiently low or the code distance is small.
At high physical error rates or large code distances, however, the alternating ordering again has the lower logical error rate. We attribute this crossover to the same mechanism discussed for the memory experiments. Notably, the regime in which the ZX interleaving ordering outperforms the alternating ordering is broader here than in the memory experiments. As expected, the N/Z ordering yields the worst performance, consistent with its fault distance of $d/2$.

\section{Conclusion}
\label{sec:conclusion}
In this work, we proposed the ZX interleaving syndrome extraction, a new method that preserves the fault distance of the surface code for arbitrary layouts with regular stabilizer tiles at minimum depth. Unlike the existing approaches in Refs.~\cite{kishony2026surface,litinski2018lattice}, the proposed method requires no additional circuit depth, imposes no non-uniformity of CNOT orderings across the surface code, and avoids the simultaneous execution of measurement or reset operations alongside CNOT gates. Along these lines, it is important that our numerical simulations under uniform depolarizing noise confirmed that the proposed method achieves the full fault distance for both memory experiment and lattice-surgery operation, and that the logical error rate is lowest among existing methods when the physical error rate is sufficiently low. The ability to construct distance-preserving syndrome extraction circuits for any surface-code layout at minimum depth, without imposing additional hardware constraints and even with relaxing the connectivity requirements, paves the way for performing universal logical operations on the surface code under realistic device conditions. We believe this technique is indispensable for the practical realization of fault-tolerant quantum computers.

Several directions for future work remain.
First, at higher physical error rates or larger code distances, the proposed method exhibits a higher logical error rate than the existing methods, owing to the additional measurement qubits and the extra error events that they introduce. Further investigation into modified constructions for syndrome-extraction circuits that mitigate these additional error events is needed.
Second, the method can be applied to the layouts considered in Refs.~\cite{fujiu2025dense,gidney2025factor,gidney2025yoked} while preserving the fault distance at minimum depth. We believe that logical error rates in these layouts will also improve when physical error rates are sufficiently small. 
Third, since the ZX interleaving scheme eliminates hook errors, it opens the possibility of designing new protocols for logical Clifford gates, such as H and S, on the surface code with a smaller spacetime volume than the existing constructions. Applying the proposed method to irregular stabilizers such as twist defects appearing in the bulk of the surface code, in Ref.~\cite{litinski2018lattice}, for example, is also a vital direction.

Lastly, some readers may wonder whether extending the alternating ordering by introducing additional qubits, similar to the ZX interleaving ordering, would provide a better way to achieve the full fault distance. A straightforward approach would be to enlarge the layout by one unit to realize fault distance $d$. Nevertheless, our proposal has an advantage compared with this approach because such an approach reduces the logical-qubit density per unit area compared to the proposed method. In other words, our method uses only the nearest-unused-measurement qubits, whereas the alternative approach would require not only the nearest measurement qubits but also the nearest-unused-data qubits, thereby lowering the density of logical qubits on a hexagonal or square grid. It is also conceivable that a more refined approach could use the same number of additional qubits as the proposed method and employ them as flag qubits to detect error events that significantly contribute to the logical error rate in the alternating ordering. We agree that such an approach may ultimately prove superior; however, we do not currently know how to construct it explicitly. Therefore, developing such a method remains an important direction for future work.

\section*{Acknowledgments}
YH proposed the original idea, conducted the numerical calculations, and prepared the manuscript. 
SI contributed to discussions on the optimization of syndrome-extraction circuits. 
YU and YS supervised the project. 
All authors reviewed, revised, and approved the final manuscript. 
This work is supported by MEXT Q-LEAP Grant No.~JPMXS0120319794 and JPMXS0118068682, JST Moonshot R\&D Grant No.~JPMJMS2061, JST CREST Grant No.~JPMJCR23I4, JPMJCR24I4, and JPMJCR25I4, JSPS KAKENHI Grant No.~JP22H05000 and JP25K21176, and RIKEN Special Postdoctoral Researcher Program.

\clearpage

\bibliography{ref}

\end{document}